# RMT Estimator with Adaptive Decision Criteria for Estimating the Number of Signals Based on Random Matrix Theory

Huiyue Yi, *Member, IEEE*

***Abstract*—Estimating the number of signals embedded in noise is a fundamental problem in signal processing. As a classic estimator based on random matrix theory (RMT), the RMT estimator estimates the number of signals via sequentially testing the likelihood of an eigenvalue as arising from a signal or noise for a given over-detection probability. However, it tends to under-estimate the number of signals as weak signal eigenvalues may be immersed in the non-negligible bias term among eigenvalues for finite sample size. In order to overcome the higher under-estimation probability of the RMT estimator, we propose an RMT estimator with adaptive decision criterion (termed as "RMT-ADC estimator") by adaptively incorporating the bias term into the decision criterion of the RMT estimator. Firstly, we analyze the effect of this bias term among eigenvalues on the estimation performance of the RMT estimator. Then, we derive both the decreased over-estimation probability and the increased under-estimation probability of the RMT estimator incurred by the bias term when assuming the eigenvalue being tested is arising from a signal, and also derive the increased under-estimation probability of the RMT estimator incurred by the bias term when assuming the eigenvalue being tested is arising from noise. Based on these results, the RMT-ADC estimator can adaptively determine whether the noise variance should be estimated under the assumption that the eigenvalue being tested is arising from a signal or from noise, and thus can adaptively select its decision criterion. Moreover, the RMT-ADC estimator can adaptively determine whether the bias term among eigenvalues should be incorporated into the selected decision criterion or not. Therefore, the RMT-ADC estimator can successfully detect signal eigenvalues immersed in this bias term, and thus avoids the higher under-estimation probability of the RMT estimator. In addition, the RMT-ADC estimator can also avoid the higher over-estimation probability of the RMT estimator resulted from the under-estimated noise variance. Finally, simulation results are presented to show that the proposed RMT-ADC estimator significantly outperforms the existing estimators.**

***Index Terms*—Detection and estimation, random matrix theory, sample covariance matrix, Lawley's law, signal number estimation, sample eigenvalue, Tracy-Widom distribution.**

This work was supported by the National Key project under Grant 2018ZX03001030, and Shanghai Municipal Natural Science Foundation under Grant 18ZR1437600.

Huiyue Yi is with the Shanghai Research Center for Wireless Communications (WiCO) and Shanghai Institute of Microsystem and information technology, Chinese Academy of Science, 3/F, Xinwei Building A, 1455 Pingcheng Road, Jiading, Shanghai 201800, P. R. China (e-mail: huiyue.yi@mail.sim.ac.cn; huiyue_yi@263.net).

## I. Introduction

ESTIMATING the number of signals in a linear mixture model is a fundamental problem in statistical signal processing and array signal processing [1]-[6]. In the signal processing literature, two most common estimators for this problem are the Akaike information criterion (AIC) and minimum description length (MDL) [7]-[8] which are based on the eigenvalues of the sample covariance matrix [7]. As noted in [24]-[25], though the MDL estimator is consistent as sample size $n \to \infty$, it fails to detect signals at low signal-to-noise ratio (SNR). In contrast, while the AIC estimator is able to detect low SNR signals, it is inconsistent as $n \to \infty$, having a non-negligible probability to overestimate the number of signals for $n >> 1$. Moreover, neither of MDL and AIC estimators performs well when the system size is comparable to the sample size. In addition, neither of MDL and AIC is applicable to large aperture arrays with a large number of sensors larger than the number of samples.

The large random matrix theory [10]-[11] has become a powerful tool to deal with the case when the sample size is of the same order of the system size. The random matrix theory concerns both the distribution of noise eigenvalues and of the signal eigenvalues in the large-system-size large-sample-size asymptotic region [12]-[22]. As is justified by these works, the random matrix theory provides a more precise approximation for the distribution of the sample eigenvalues in finite sample size settings than the classical multivariate statistical theory. In recent years, the use of random matrix theory in estimating the number of signals or weak signal detection has attracted much attention [5], [23]-[28]. In these methods, results on the spectral behavior of random matrices are applied to the problem of detecting the number of signals in a noisy linear mixture. As shown in [12]-[15], the fluctuation of the largest noise eigenvalue of the sample covariance matrix can be modeled by the celebrated Tracy-Widom distribution under the assumption of Gaussian data. Based on this result, the authors in [24] proposed a RMT estimator to estimate the number of signals via detecting the largest noise eigenvalues as arising from a signal or noise for a given over-detection probability. In this RMT estimator, a method is provided to estimate the noise level, and the Tracy-Widom distribution is utilized to construct the thresholds for the sequential tests. In the sequential tests, the thresholds are designed to control the overestimation probability. In [5], a two-step test procedure based on random matrix theory is proposed for source enumeration. In this method, the second step is based on a likelihood ratio test to reduce the underestimation occurred in its first-step test. As illustrated in [5], this second-step test is suboptimal because only the marginal PDFs are utilized to compute the likelihood ratio, and it is not easy to derive an explicit expression for the test threshold. In [25], the authors analyze the detection performance of the AIC estimator from the random matrix theory point, and propose a modified AIC estimator with a small increase in the penalty term. This modified AIC estimator has a better detection performance than the MDL with a negligible overestimation probability. Moreover, finding the optimal penalty term for the AIC and MDL is still an open question. As were analyzed in [24]-[28], a shortcoming of the sample-eigenvalue-based detection schemes is that it just might not be possible to detect low-level or closely spaced signals when there are too few samples available. In other words, if the signals are not strong enough and not spaced far enough part, then not only will the RMT estimator consistently under-estimate the number of signals but so will any other sample-eigenvalue-based detectors.

As proved in [29]-[31], there exists a non-negligible bias term among eigenvalues when the number of samples is limited. However, as illustrated in [24], the RMT estimator does not consider the bias term among eigenvalues when the sample size is limited. Since weak or even very strong signal eigenvalues will be immersed in this bias term, the RMT estimator tends to under-estimate the number of signals when the sample size is limited. To our best knowledge, the impact of this bias term among eigenvalues on the detection performance of the existing signal number estimators is still an open problem to be solved.

In order to solve the above problem, in this paper we propose an RMT estimator with adaptive decision criterion (termed as “RMT-ADC estimator”) by adaptively incorporating the bias term into the decision criterion of the RMT estimator. In the development of the RMT-ADC estimator, we utilize the results regarding the asymptotically norm distribution of the sample signal eigenvalues [18]-[21] and its expectations [29]-[31], which reflects the interaction among eigenvalues for limited sample size and system size. The proposed RMT-ADC estimator can adaptively determine whether the noise variance should be estimated under the assumption that the eigenvalue being tested is arising from a signal or from noise, can adaptively select the decision criterion, and can determine whether the bias term should be incorporated into the selected decision criterion or not. Therefore, the RMT-ADC estimator can successfully detect the signal eigenvalues immersed in the bias term among eigenvalues. The main contributions of this work are summarized as follows:

(1) In Section Ⅲ. A, we analyze the effect of the bias term among eigenvalues on the estimation performance of the RMT estimator. Specifically, we derive the analytical formulas for both the increased under-estimation probability and the decreased over-estimation probability of the RMT estimator incurred by the bias term among eigenvalues.

(2) In Section Ⅲ. B, we firstly derive both the decreased over-estimation probability and the increased under-estimation probability of the RMT estimator incurred by the bias term among eigenvalues when assuming that the eigenvalue being tested is arising from a signal. Secondly, we derive the increased under-estimation probability of the RMT estimator incurred by the bias term among eigenvalues when assuming the eigenvalue being tested is arising from noise. Based on these results, the proposed RMT-ADC estimator can adaptively determine whether the noise variance should be estimated under the assumption that the eigenvalue being tested is arising from a signal or from noise, and thus can adaptively select its decision criterion. Moreover, the RMT-ADC estimator can adaptively determine whether the bias term among eigenvalues should be incorporated into the selected decision criterion or not. Therefore, the RMT-ADC estimator can successfully detect signal eigenvalues immersed in this bias term, and thus avoids the higher under-estimation probability of the RMT estimator. In addition, the RMT-ADC estimator can also overcome the higher over-estimation probability of the RMT estimator resulted from the under-estimated noise variance.

(3) Extensive simulations are presented to compare the estimation performance of the proposed RMT-ADC estimator to the existing methods including the RMT estimator [24], the classic AIC and MDL estimator [7]-[8], and the modified AIC estimator [25]. Simulation results show that the proposed RMT-ADC estimator can overcome the higher down-estimation probability of the RMT estimator. Moreover, the over-estimation probability of the proposed RMT-ADC estimator can be controlled by the pre-fixed false alarm probability. In addition, the proposed RMT-ADC estimator significantly outperforms the existing estimators.

This paper is organized as follows. In Section II, we present the problem formulation, the mathematical preliminaries from the random matrix theory and the prior works. In Section III, we firstly analyze the effect of the bias term among eigenvalues on the estimation performance of the RMT estimator, and describe the proposed RMT-ADC estimator. Simulation results that illustrate the superior estimation performance of the proposed RMT-ADC estimator over the existing methods are presented in Section IV. Finally, conclusions are drawn in Section V.

## II. Problem Formulation, Random Matrix Theory And Prior Works

In this section, we firstly introduce the data model and problem formulation. Then, we provide the mathematical preliminaries from the random matrix theory and the Lawley’s law. Finally, we describe the RMT estimator in [24], which will be utilized in the development of our RMT-ADC estimator in Section III.

### A. Data Model and Problem Formulation

We consider the following standard model $p$ -dimensional linear mixture model for signals impinging on an array with $p$

sensors [24]. Let $\{\mathbf{x}(i)=\mathbf{x}(t_i)\}_{i=1}^{n}$ denote $n$ i.i.d. observations of the form

$$\mathbf{x}(t)=\sum_{i=1}^{q}\mathbf{v}_i s_i(t)+\mathbf{w}(t)=\mathbf{A}\mathbf{s}(t)+\mathbf{w}(t), \qquad (1)$$

sampled at distinct times $t_i$, where $\mathbf{s}(t)=[s_1(t),\cdots,s_q(t)]^T$ is a $q\times 1$ vector containing $q$ different zero-mean signal components with corresponding independent array response vectors $\mathbf{v}_i\in\mathrm{R}^p$, $\mathbf{A}=[\mathbf{v}_1,\mathbf{v}_2,\cdots,\mathbf{v}_q]$ is the array response matrix, and the noise $\mathbf{w}(t)\in\mathrm{R}^p$ are assumed to be additive white Gaussian noise (AWGN) with zero mean and unknown variance $\sigma^2$, i.e., $\mathbf{w}(t)\sim N(\mathbf{0},\sigma^2\mathbf{I}_p)$, and $\mathbf{w}(t)$ is independent of $\mathbf{s}(t)$. In addition, we assume the $q\times q$ covariance matrix $\mathbf{\Sigma}_s=E[\mathbf{s}\mathbf{s}^H]$ is of full rank. Under these assumptions, the population covariance matrix of the observations $\mathbf{x}$ is given by $\Sigma=E[\mathbf{x}\mathbf{x}^H]$ with its $q$ noise-free population signal eigenvalues given by $\{\lambda_1,\lambda_2,\cdots,\lambda_q\}$, and thus the population eigenvalues of $\mathbf{\Sigma}$ is given by

$$\{\lambda_1+\sigma^2,\cdots,\lambda_q+\sigma^2,\sigma^2,\cdots,\sigma^2\}. \qquad (2)$$

Then, if the true covariance matrix $\mathbf{\Sigma}$ was known, the dimension of the signal dimensions can be determined from the smallest eigenvalues of $\mathbf{\Sigma}$. In practice, the problem is that we can only get finite samples of observations and thus the true covariance matrix $\mathbf{\Sigma}$ is unknown. As a result, the problem is to determine the number $q$ of signal components from $n$ finite i.i.d. noisy samples $\{\mathbf{x}(i)\}_{i=1}^{n}$ of $p$ -dimensional real or complex Gaussian snapshot vectors in (1).

We denote by $\mathbf{S}_n$ the sample covariance matrix of the $n$ samples $\{\mathbf{x}(i)\}_{i=1}^{n}$ from the model (1),

$$\mathbf{S}_n=\frac{1}{n}\sum_{i=1}^{n}\mathbf{x}(i)\mathbf{x}^H(i). \qquad (3)$$

Let the sample eigenvalues of $\mathbf{S}_n$ be $l_1\geq l_2\geq\cdots\geq l_p$. Estimating the number of signals $q$ from finite samples is a model order selection problem for which there are many approaches. In the nonparametric setting, most methods are based on the eigenvalues of the sample covariance matrix. In particular, two well-known classical AIC and MDL estimators [7]-[8] are based on the fact that the sample covariance approximates the population covariance matrix well when sample size is large. However, this does not hold for the case when $p/n\to\gamma\in(0,\infty)$ as $n\to\infty$.

The random matrix theory is a powerful tool to characterize the distribution of the sample eigenvalues for the case when $p/n\to\gamma\in(0,\infty)$ as $n\to\infty$ [10]-[22]. Nevertheless, the random matrix theory has been used for signal detection and estimation [5]-[6], [23]-[28], and these methods have superior detection performance over the classical methods. In this paper, we will further consider inferring the unknown number $q$ of signals from the $n$ samples $\{\mathbf{x}(i)\}_{i=1}^{n}$ under the nonparametric setting from the viewpoint of random matrix theory.

*B. Mathematical Preliminaries: Random Matrix Theory and Lawley's Law*

In most cases, the number of sources is much smaller than the system size, i.e., $q<<p$, which means that the population covariance matrix $\Sigma=E[\mathbf{x}\mathbf{x}^H]$ is a low rank perturbation of an identity matrix. Such a population covariance matrix is called as the spiked covariance model [16]-[22], where all eigenvalues of the population covariance matrix are equal except for a small fixed number of distinct "spike eigenvalues". As the key goal in nonparametric estimation of the number of sources is to distinguish between noise and signal eigenvalues, in this subsection we will review some related results under this spiked covariance model regarding the asymptotic distribution of the eigenvalues of the sample covariance matrix $\mathbf{S}_n$.

The first result describes the asymptotic distribution of the largest eigenvalue of a pure noise matrix [12]-[15]. Let $\mathbf{S}_n$ denote the sample covariance matrix of pure noise observations distributed as $N(\mathbf{0},\sigma^2\mathbf{I}_p)$. In the joint limit $p,n\to\infty$ with $p/n\to\gamma\in(0,\infty)$, the distribution of the largest eigenvalue of $\mathbf{S}_n$ converges to the Tracy-Widom distribution. That is, for every $x\in\mathrm{R}$,

$$\Pr[l_1<\sigma^2(\mu_{n,p}+x\sigma_{n,p})]\to F_\beta(x). \qquad (4)$$

where $\beta=1$ for real valued noise and $\beta=2$ for complex-valued noise. The centering and scaling parameters $\mu_{n,p}$ and $\sigma_{n,p}$, respectively, are functions of $n$ and $p$ only [12]-[15]. For real valued noise, the following formulas provide $O(p^{-2/3})$ convergence rate in (4), see [14]

$$\mu_{n,p}=\frac{1}{n}(\sqrt{n-1/2}+\sqrt{p-1/2})^2, \qquad (5)$$

$$\sigma_{n,p}=\sqrt{\frac{\mu_{n,p}}{n}}\left(\frac{1}{\sqrt{n-1/2}}+\frac{1}{\sqrt{p-1/2}}\right)^{1/3}. \qquad (6)$$

The second result describes the phase transition phenomenon for the signal eigenvalues in the spiked covariance model [16]-[20]. If the signal strength is not larger than a certain threshold, the corresponding signal eigenvalue converges to the upper limit of the support of the Marcenko-Pastur density, otherwise it is pulled up to a higher limit. Suppose that the fourth moment of the entries of $\mathbf{S}_n$ exists. Then, in the joint limit $p,n\to\infty$, with

$p/n \to \gamma \in (0,\infty)$, the $i$ th signal sample eigenvalue $l_i$ converges with probability one to

$$l_i \xrightarrow{\text{a.s.}} \begin{cases} (\lambda_i+\sigma^2)(1+\gamma\sigma^2/\lambda_i), & \text{if } \lambda_i > \sigma^2\sqrt{\gamma} \\ \sigma^2(1+\sqrt{\gamma})^2, & \text{if } \lambda_i \le \sigma^2\sqrt{\gamma} \end{cases}, \quad i = 1, 2, \cdots, q, \tag{7}$$

where the threshold $\sigma^2\sqrt{\gamma}$ is called as the non-parametric asymptotic limit of detection, which can be denoted as

$$\lambda_{\text{DET}} = \sigma^2\sqrt{\gamma}. \tag{8}$$

This detection threshold captures the fundamental limit of the sample eigenvalue-based source number estimation methods in [5], [23]-[28], which means that the asymptotically detectable signal must have signal strength larger than $\lambda_{\text{DET}}$.

The third result characterizes the limiting distributions of the signal eigenvalues with strength $\lambda_i > \sigma^2\sqrt{\gamma}$ [18]-[21]. Such signal eigenvalues are distributed normally around the limiting value $(\lambda_i+\sigma^2)(1+\gamma\sigma^2/\lambda_i)$ given in (7). For the $q$ th signal with $\lambda_q > \sigma^2\sqrt{\gamma}$, in the joint limit $p, n \to \infty$, with $p/n \to \gamma \in (0,\infty)$, at a convergence rate of $O(n^{-1/2})$, the fluctuations of $l_q$ converges with probability one to the normal distribution:

$$l_q \xrightarrow{D} N(\tau(\lambda_q), \delta^2(\lambda_q)), \tag{9}$$

where $\xrightarrow{D}$ denotes convergence in distribution, and

$$\tau(\lambda) = (\lambda+\sigma^2)\left(1+\frac{p-q}{n}\cdot\frac{\sigma^2}{\lambda}\right), \tag{10}$$

$$\delta(\lambda) = (\lambda+\sigma^2)\sqrt{\frac{2}{\beta n}\left(1-\frac{p-q}{n}\cdot\frac{\sigma^4}{\lambda^2}\right)}. \tag{11}$$

While asymptotically there are no signal-signal interactions among signal eigenvalues, we need to take into account the non-negligible interaction among eigenvalues for finite $p$ and $n$. In [29]-[31], a more accurate expression for the expectation value of the sample eigenvalue $l_j$ for $j \le q$ in the non-asymptotic region is given by:

$$E[l_j] = \rho_j + \frac{(p-q)\rho_j\sigma^2}{n(\rho_j-\sigma^2)} + \frac{\rho_j}{n}\sum_{i=1,i\ne j}^{q}\frac{\rho_i}{\rho_j-\rho_i} + O(n^{-2}) \tag{12}$$

where $\rho_j = \lambda_j + \sigma^2$. As can be seen from (12), the sample eigenvalues are highly affected by a bias term for finite $n$, which is caused by the interaction among the eigenvalues. Moreover, this bias term are non-negligible for finite values of $n$. For the sake of simplification, we define the following notations:

$$\nu_j = \frac{\lambda_j+\sigma^2}{n}\sum_{i=1,i\ne j}^{q}\frac{\lambda_i+\sigma^2}{\lambda_j-\lambda_i}, \tag{13}$$

$$\kappa_j = 1+\frac{(p-q)\sigma^2}{n\lambda_j}. \tag{14}$$

## C. Prior Works

As stated in (4), the fluctuation of the largest noise eigenvalue of the sample covariance matrix can be modeled by the Tracy-Widom distribution under the assumption of Gaussian data. Consequently, if the noise variance $\sigma^2$ is known, a statistical procedure to distinguish a signal eigenvalue $l$ from noise at a significant level $\alpha$ is to check whether $l > \sigma^2(\mu_{n,p} + s(\alpha)\sigma_{n,p})$, where the value of $s(\alpha)$ depends on the required significant level $\alpha$. Based on this observation, a RMT estimator was proposed in [24] to estimate the number of signals via detecting the largest noise eigenvalues.

The RMT estimator is based on a sequence of hypothesis tests, for $k = 1, 2, \cdots, \min(p,n)-1$,

$$\begin{aligned} H_k &: \text{ at least } k \text{ components}, \\ H_{k-1} &: \text{ at most } k-1 \text{ components}. \end{aligned} \tag{15}$$

For each value of $k$, the noise level $\sigma^2$ and the signal eigenvalue $\{\lambda_i\}_{i=1}^{k}$ are estimated assuming that $l_{k+1}, \cdots, l_p$ correspond to noise via solutions of the following non-linear system of equations [24]:

$$\sigma^2_{\text{RMT}}(k) - \frac{1}{p-k}\left[\sum_{j=k+1}^{p} l_j + \sum_{j=1}^{k}(l_j - \hat{\rho}_j)\right] = 0, \tag{16}$$

$$\hat{\rho}_j^2 - \hat{\rho}_j\left[l_j + (1-\frac{p-k}{n})\sigma^2_{\text{RMT}}(k)\right] + l_j\sigma^2_{\text{RMT}}(k) = 0. \tag{17}$$

This system is solved iteratively starting from an initial value $\hat{\sigma}_0^2$ given by its maximum likelihood estimate $\hat{\sigma}_0^2 = 1/(p-k)\cdot\sum_{j=k+1}^{p} l_j$. After the convergence of the above system, we obtain the estimates for $\{\hat{\rho}_i\}_{i=1}^{k}$ and noise level $\sigma^2_{\text{RMT}}(k)$. Then, the signal eigenvalue $\lambda_i$ is estimated as $\hat{\lambda}_i = \hat{\rho}_i - \sigma^2_{\text{RMT}}(k)$.

Then, the likelihood of the $k$ th eigenvalue $l_k$ is tested as arising from a signal or from noise as follows:

$$l_k > \sigma^2_{\text{RMT}}(k)\left(\mu_{n,p-k} + s(\alpha)\sigma_{n,p-k}\right). \tag{18}$$

For this test to have a false alarm with asymptotic probability $\alpha$ as $p, n \to \infty$, the threshold $s(\alpha)$ should satisfy

$$F_\beta(s(\alpha)) = 1-\alpha. \tag{19}$$

The value of $s(\alpha)$ can be calculated by inverting the Tracy-Widom distribution, and this inversion can be computed numerically by using the software package [32]. If (18) is satisfied, $H_k$ is accepted and $k$ is increased by one. Otherwise, $\hat{q} = k - 1$. That is to say

$$\hat{q} = \arg\min_k \left\{ l_k < \sigma^2_{\mathrm{RMT}}(k)\left(\mu_{n,p-k} + s(\alpha)\sigma_{n,p-k}\right)\right\} - 1. \tag{20}$$

As can be seen from (12), there exists a non-negligible bias term among eigenvalues when the number of sample size $n$ is limited. However, the decision criterion in (18) of the RMT estimator does not consider this bias term among eigenvalues, and thus this bias term will affect the detection performance of the RMT estimator for finite $p$ and $n$. As will be analyzed in Section III. A, the RMT estimator tends to under-estimate the number of signals as some signal eigenvalues will be immersed in this bias term. In order to overcome this problem, in next Section we will develop an RMT-ADC estimator by utilizing the results regarding the asymptotically norm distribution of the sample signal eigenvalues given by (9) and the expectation value of the sample eigenvalues given by (12) for finite $p$ and $n$.

## III. RMT Estimator with Adaptive Decision Criterion for Signal Number Estimation

In this Section, we firstly analyze the effect of the non-negligible bias term among eigenvalues on the estimation performance of the RMT estimator [24] for finite $p$ and $n$. Then, we propose an RMT estimator with adaptive decision criterion (termed as "RMT-ADC estimator") by adaptively incorporating this bias term into the RMT estimator. Finally, we provide performance comparison of the proposed RMT-ADC estimator with the RMT estimator.

### *A. Performance Analysis of the RMT estimator*

Denote by $H_q$ the hypothesis that the true number of sources is $q$. Then, the probability of estimating the number of signals incorrectly (i.e., misestimation probability) is defined as

$$P_{\mathrm{e}} = P(\hat{q} \neq q \mid H_q) = P_{\mathrm{m}} + P_{\mathrm{f}}. \tag{21}$$

where the probability of under-estimation $P_m$ and the probability of overe-stimation $P_{\mathrm{f}}$ are, respectively, given by

$$P_{\mathrm{m}} = P(\hat{q} < q \mid H_q), \tag{22}$$

$$P_{\mathrm{f}} = P(\hat{q} > q \mid H_q). \tag{23}$$

For notational convenience, we denote the right hand side (RHS) of (18) as

$$\varphi_{n,p-q} = \sigma^2\left(\mu_{n,p-q} + s(\alpha)\sigma_{n,p-q}\right). \tag{24}$$

The condition for the RMT estimator to determine at least the correct number of signals $q$ is [24]

$$l_q > \varphi_{n,p-q}. \tag{25}$$

From (13), the bias term $v_q$ is expressed as

$$v_q = \frac{\lambda_q + \sigma^2}{n}\sum_{i=1}^{q-1}\frac{\lambda_i + \sigma^2}{\lambda_q - \lambda_i}. \tag{26}$$

Obviously, $v_q < 0$. Therefore, $v_q$ will decrease the value of the eigenvalue $l_q$, and thus the RMT estimator will under-estimate the number of signals as some eigenvalues will be immersed in the bias term $v_q$.

In order to analyze the effect of the bias term $v_q$ on the estimation performance of the RMT estimator, we incorporate the bias term $v_q$ into the RMT estimator to introduce an estimator termed as "b-RMT estimator", which tests the likelihood of the eigenvalue $l_q$ as arising from a signal or from noise according to the following decision condition:

$$l_q - v_q > \varphi_{n,p-q}. \tag{27}$$

Then, we analyze the effect of the bias term $v_q$ on the estimation performance of the RMT estimator for finite values of $p$ and $n$ according to the following two cases:

(1) In the first case, we assume that $\lambda_q$ has multiplicity one and $\sigma^2\sqrt{\gamma} < \lambda_q << \lambda_{q-1}$ such that $l_q - v_q > \varphi_{n,p-q}$ and $l_q < \varphi_{n,p-q}$. Obviously, the signal eigenvalue $l_q$ may be immersed in this bias term $v_q$, and thus can't be detected by the RMT estimator. Therefore, the main source of error is miss-detection of the $q$th sample eigenvalue $l_q$, and thus the RMT will under-estimate the number of signals. In the following, we will derive the increased under-estimation probability of the RMT estimator incurred by $v_q$.

From (12), we introduce the following statistics:

$$z_i = (l_i - v_i)/\kappa_i - \sigma^2. \tag{28}$$

According to (9), it is easy to derive that $z_i$ follows the following normal distribution:

$$z_i \xrightarrow{D} N(\tau_i,\, \omega_i^2). \tag{29}$$

From (10) and (11), the mean $\tau_i$ and standard deviation $\omega_i$ are, respectively, derived as:

$$\tau_i = E[z_i] = \frac{E[l_i]}{\kappa_i} - \frac{v_i}{\kappa_i} - \sigma^2 = \lambda_i, \tag{30}$$

$$\omega_i = \frac{\lambda_i + \sigma^2}{\kappa_i}\sqrt{\frac{2}{\beta n}(1 - \frac{p-q}{n}\cdot\frac{\sigma^4}{\lambda_i^2})}\,. \quad (31)$$

Then, we firstly derive the misdetection probability of the b-RMT estimator. According to (29), the under-detection probability of the b-RMT estimator is derived as

$$P_{\mathrm{m}}^{\text{b-RMT,s}}(q,\nu_q) = \Pr\left\{\eta < \frac{(\varphi_{n,p-q} + \nu_q)/\kappa_q - \sigma^2 - \lambda_q}{\omega_q}\right\}. \quad (32)$$

Where $\eta \sim N(0,1)$. Then, the misdetection probability of the b-RMT estimator is given by

$$P_{\mathrm{e}}^{\text{b-RMT,s}}(q,\nu_q) = P_{\mathrm{m}}^{\text{b-RMT,s}}(q,\nu_q) + \alpha\,. \quad (33)$$

Secondly, we derive the misdetection probability of the RMT estimator. Without considering $\nu_q$, the under-detection probability and over-detection probability of the RMT estimator are, respectively, given by

$$P_{\mathrm{m}}^{\text{RMT,s}}(q,\nu_q) = \Pr\left\{\eta < \frac{\varphi_{n,p-q}/\kappa_q - \sigma^2 - \lambda_q}{\omega_q}\right\}, \quad (34)$$

$$P_{\mathrm{f}}^{\text{RMT,s}}(q,\nu_q) = 1 - F_\beta\left(s(\alpha) - \frac{\nu_q}{\sigma^2\cdot\sigma_{n,p-q}}\right). \quad (35)$$

From (34) and (35), the misdetection probability of the RMT estimator is derived as

$$P_{\mathrm{e}}^{\text{RMT,s}}(q,\nu_q) = P_{\mathrm{m}}^{\text{RMT,s}}(q,\nu_q) + P_{\mathrm{f}}^{\text{RMT,s}}(q,\nu_q). \quad (36)$$

From (33) and (36), the increased under-estimation probability of the RMT estimator incurred by $\nu_q$ is derived as

$$\Delta P_{\text{UE,inc}}^{\text{RMT,s}}(q,\nu_q) = P_{\mathrm{e}}^{\text{RMT,s}}(q,\nu_q) - P_{\mathrm{e}}^{\text{b-RMT,s}}(q,\nu_q)\,. \quad \textbf{(37)}$$

(2) In the second case, we assume that $\lambda_q$ has multiplicity one and $\sigma^2\sqrt{\gamma} << \lambda_q$ such that $l_q > \varphi_{n,p-q}$. In this case, the bias term $\nu_q$ will decrease the over-estimation probability of the RMT estimator, but increase the over-estimation probability of the b-RMT estimator. In the following, we will derive the decreased over-estimation probability of the RMT estimator incurred by $\nu_q$.

When considering $\nu_q$, the over-estimation probability of the b-RMT estimator is given by

$$P_{\mathrm{f}}^{\text{b-RMT,s}}(q,\nu_q) = 1 - F_\beta\left(s(\alpha) + \frac{\nu_q}{\sigma^2\cdot\sigma_{n,p-q}}\right). \quad (38)$$

From (32) and (38), the misdetection probability of the b-RMT estimator is given by

$$P_{\mathrm{O,e}}^{\text{b-RMT,s}}(q,\nu_q) = P_{\mathrm{m}}^{\text{b-RMT,s}}(q,\nu_q) + P_{\mathrm{f}}^{\text{b-RMT,s}}(q,\nu_q)\,. \quad (39)$$

Then, the decreased over-estimation probability of the RMT estimator incurred by $\nu_q$ can be derived as

$$\Delta P_{\text{OE,dec}}^{\text{RMT,s}}(q,\nu_q) = P_{\mathrm{O,e}}^{\text{b-RMT,s}}(q,\nu_q) - P_{\mathrm{O,e}}^{\text{b-RMT,s}}(q,\nu_q)|_{\nu_q=0}\,. \quad (40)$$

From (37) and (40), the overall increased under-estimation probability of the RMT estimator incurred by $\nu_q$ is derived as

$$\Delta P_{\text{O-UE,inc}}^{\text{RMT,s}}(q,\nu_q) = \Delta P_{\text{UE,inc}}^{\text{RMT,s}}(q,\nu_q) - \Delta P_{\text{OE,dec}}^{\text{RMT,s}}(q,\nu_q)\,. \quad (41)$$

As can be seen from the above analysis, the RMT estimator has higher under-estimation probability as some signal eigenvalues will be immersed in the bias term among eigenvalues for finite sample size $n$. This observation motivates us to develop an RMT estimator with adaptive decision criterion in next subsection by adaptively incorporating the bias term among eigenvalues into the RMT estimator.

### *B. RMT Estimator with Adaptive Decision Criterion: RMT-ADC estimator*

In order to overcome the higher under-estimation probability of the RMT estimator in (18), we propose a RMT-ADC estimator by adaptively incorporating the bias term into the RMT estimator. As in (15), the RMT-ADC estimator is based on a sequence of hypothesis tests. For $k = 1, 2, \cdots, \min(p,n) - 1$, it tests the likelihood of the $k$th eigenvalue $l_k$ as arising from a signal or from noise according to the following four steps.

**Step 1:**

We derive the overall increased under-estimation probability of the RMT estimator incurred by $\hat{\nu}_k$ under the assumption that $l_k$ is arising from a signal.

*(1) Increased under-estimation probability of the RMT estimator incurred by $\hat{\nu}_k$ under the assumption that $l_k$ is arising from a signal*

Firstly, we derive the misdetection probability of the b-RMT estimator under the assumption that $l_k$ is arising from a signal.

We denote the right hand side of (18) as

$$\hat{\varphi}_{n,p-k} = \sigma_{\mathrm{RMT}}^2(k)\left(\mu_{n,p-k} + s(\alpha)\sigma_{n,p-k}\right). \quad (42)$$

Then, the decision condition of the b-RMT estimator in (27) is rewritten as

$$l_k - \hat{\nu}_k > \hat{\varphi}_{n,p-k}\,. \quad (43)$$

Similarly to (32), the under-estimation probability of the b-RMT estimator is given by

$$\hat{P}_{\mathrm{m}}^{\text{b-RMT,s}}(k,\hat{\nu}_k) = \Pr\left\{\eta < \frac{(\hat{\varphi}_{n,p-k} + \hat{\nu}_k)/\hat{\kappa}_k - \sigma_{\mathrm{RMT}}^2(k) - \hat{\lambda}_k}{\hat{\omega}_k}\right\}. \quad (44)$$

Then, the misdetection probability of the b-RMT estimator under the assumption that $l_k$ is arising from a signal is given by

$$\hat{P}_{\mathrm{e}}^{\text{b-RMT,s}}(k,\hat{v}_k)=\hat{P}_{\mathrm{m}}^{\text{b-RMT,s}}(k,\hat{v}_k)+\alpha. \quad (45)$$

Secondly, we derive the misdetection probability of the RMT estimator. Similarly to (34) and (35), the under-detection probability and over-detection probability of the RMT estimator are, respectively, given by

$$\hat{P}_{\mathrm{m}}^{\text{RMT,s}}(k,\hat{v}_k)=\Pr\left\{\eta<\frac{\hat{\varphi}_{n,p-k}/\hat{\kappa}_k-\sigma_{\mathrm{RMT}}^2(k)-\hat{\lambda}_k}{\hat{\omega}_k}\right\}, \quad (46)$$

$$\hat{P}_{\mathrm{f}}^{\text{RMT,s}}(k,\hat{v}_k)=1-F_\beta\left(s(\alpha)-\frac{\hat{v}_k}{\sigma_{\mathrm{RMT}}^2(k)\cdot\sigma_{n,p-k}}\right). \quad (47)$$

From (46) and (47), the misdetection probability of the RMT estimator is given by

$$\hat{P}_{\mathrm{e}}^{\text{RMT,s}}(k,\hat{v}_k)=\hat{P}_{\mathrm{m}}^{\text{RMT,s}}(k,\hat{v}_k)+\hat{P}_{\mathrm{f}}^{\text{RMT,s}}(k,\hat{v}_k). \quad (48)$$

From (45) and (48), the increased under-estimation probability of the RMT estimator incurred by $\hat{v}_k$ is given by

$$\Delta\hat{P}_{\text{UE,inc}}^{\text{RMT,s}}(k,\hat{v}_k)=\hat{P}_{\mathrm{e}}^{\text{RMT,s}}(k,\hat{v}_k)-\hat{P}_{\mathrm{e}}^{\text{b-RMT,s}}(k,\hat{v}_k). \quad (49)$$

*(2) Decreased over-estimation probability of the RMT estimator incurred by $\hat{v}_k$ under the assumption that $l_k$ is arising from a signal*

Similarly to (38), the over-estimation probability of the b-RMT estimator is given by

$$\hat{P}_{\mathrm{f}}^{\text{b-RMT,}s}(k,\hat{v}_k)=1-F_\beta\left(s(\alpha)+\frac{\hat{v}_k}{\sigma_{\mathrm{RMT}}^2(k)\cdot\sigma_{n,p-k}}\right). \quad (50)$$

From (44) and (50), the misdetection probability of the b-RMT estimator is given by

$$\hat{P}_{\mathrm{O,e}}^{\text{b-RMT,}s}(k,\hat{v}_k)=\hat{P}_{\mathrm{m}}^{\text{b-RMT,s}}(k,\hat{v}_k)+\hat{P}_{\mathrm{f}}^{\text{b-RMT,s}}(k,\hat{v}_k). \quad (51)$$

Then, the decreased over-estimation probability of the RMT estimator incurred by $\hat{v}_k$ is derived as

$$\Delta\hat{P}_{\text{OE, dec}}^{\text{RMT,s}}(k,\hat{v}_k)=\hat{P}_{\mathrm{O,e}}^{\text{b-RMT,}s}(k,\hat{v}_k)-\hat{P}_{\mathrm{O,e}}^{\text{b-RMT,}s}(k,\hat{v}_k)\big|_{\hat{v}_k=0}. \quad (52)$$

According to (49) and (52), the overall increased under-estimation probability of the RMT estimator incurred by $\hat{v}_k$ under the assumption that $l_k$ is arising from a signal is derived as

$$\Delta\hat{P}_{\text{O-UE, inc}}^{\text{RMT,s}}(k,\hat{v}_k)=\Delta\hat{P}_{\text{UE,inc}}^{\text{RMT,s}}(k,\hat{v}_k)-\Delta\hat{P}_{\text{OE, dec}}^{\text{RMT,s}}(k,\hat{v}_k). \quad (53)$$

**Step 2:**

We derive the overall increased under-estimation probability of the RMT estimator incurred by $\hat{v}_k$ under the assumption that $l_k$ is arising from noise.

When assuming that $l_k$ is arising from noise, the noise level in (18) should be estimated as $\sigma_{\mathrm{RMT}}^2(k-1)$ assuming that $l_k$, ... , $l_p$ correspond to noise. Then, the decision condition for the RMT estimator in (18) is modified as

$$l_k>\hat{\varphi}_{n,p-(k-1)}. \quad (54)$$

In addition, the decision condition for the b-RMT estimator in (43) is modified as

$$l_k-\hat{v}_k>\hat{\varphi}_{n,p-(k-1)}. \quad (55)$$

*1) Increased under-estimation probability of the RMT estimator incurred by $\hat{v}_k$ under the assumption that $l_k$ is arising from noise*

Firstly, we derive the misdetection probability of the b-RMT estimator. Similarly to (44), the under-estimation probability of the b-RMT estimator is given by

$$\hat{P}_{\mathrm{m}}^{\text{b-RMT,}n}(k,\hat{v}_k)=\Pr\left\{\eta<\frac{(\hat{\varphi}_{n,p-(k-1)}+\hat{v}_k)/\hat{\kappa}_k-\sigma_{\mathrm{RMT}}^2(k-1)-\hat{\lambda}_k}{\hat{\omega}_k}\right\}. \quad (56)$$

Then, the misdetection probability of the b-RMT estimator is given by

$$\hat{P}_{\mathrm{e}}^{\text{b-RMT,}n}(k,\hat{v}_k)=\hat{P}_{\mathrm{m}}^{\text{b-RMT,}n}(k,\hat{v}_k)+\alpha. \quad (57)$$

Secondly, we derive the misdetection probability of the RMT estimator. Similarly to (46) and (47), the under-detection probability and over-detection probability of the RMT estimator are, respectively, given by

$$\hat{P}_{\mathrm{m}}^{\text{RMT,}n}(k,\hat{v}_k)=\Pr\left\{\eta<\frac{\hat{\varphi}_{n,p-(k-1)}/\hat{\kappa}_k-\sigma_{\mathrm{RMT}}^2(k-1)-\hat{\lambda}_k}{\hat{\omega}_k}\right\}, \quad (58)$$

$$\hat{P}_{\mathrm{f}}^{\text{RMT,}n}(k,\hat{v}_k)=1-F_\beta\left(s(\alpha)-\frac{\hat{v}_k}{\sigma_{\mathrm{RMT}}^2(k-1)\cdot\sigma_{n,p-(k-1)}}\right). \quad (59)$$

From (58) and (59), the misdetection probability of the RMT estimator is given by

$$\hat{P}_{\mathrm{e}}^{\text{RMT,}n}(k,\hat{v}_k)=\hat{P}_{\mathrm{m}}^{\text{RMT,}n}(k,\hat{v}_k)+\hat{P}_{\mathrm{f}}^{\text{RMT,}n}(k,\hat{v}_k). \quad (60)$$

From (57) and (60), the increased under-estimation probability of the RMT estimator incurred by $\hat{v}_k$ is given by

$$\Delta\hat{P}_{\text{UE,inc}}^{\text{RMT,n}}(k,\hat{v}_k)=\hat{P}_{\mathrm{e}}^{\text{RMT,}n}(k,\hat{v}_k)-\hat{P}_{\mathrm{e}}^{\text{b-RMT,}n}(k,\hat{v}_k). \quad (61)$$

*2) Decreased over-estimation probability of the RMT estimator incurred by $\hat{v}_k$ under the assumption that $l_k$ is arising from noise*

Similarly to (50), the over-detection probability of the b-RMT estimator is given by

$$\hat{P}_{\mathrm{f}}^{\text{b-RMT},n}(k,\hat{v}_k)=1-F_\beta\left(s(\alpha)+\frac{\hat{v}_k}{\sigma_{\mathrm{RMT}}^2(k-1)\cdot\sigma_{n,p-(k-1)}}\right)$$
(62)

From (56) and (62), the misdetection probability of the b-RMT estimator is given by

$$\hat{P}_{\mathrm{O,e}}^{\text{b-RMT},n}(k,\hat{v}_k)=\hat{P}_{\mathrm{m}}^{\text{b-RMT},n}(k,\hat{v}_k)+\hat{P}_{\mathrm{f}}^{\text{b-RMT},n}(k,\hat{v}_k).$$
(63)

Then, the decreased over-estimation probability of the RMT estimator incurred by $\hat{v}_k$ is derived as

$$\Delta\hat{P}_{\mathrm{OE,dec}}^{\mathrm{RMT,n}}(k,\hat{v}_k)=\hat{P}_{\mathrm{O,e}}^{\text{b-RMT},n}(k,\hat{v}_k)-\hat{P}_{\mathrm{O,e}}^{\text{b-RMT},n}(k,\hat{v}_k)\,|_{\hat{v}_k=0}\,.$$
(64)

According to (61) and (64), the overall increased under-estimation probability of the RMT estimator incurred by $\hat{v}_k$ under the assumption that $l_k$ is arising from noise is derived as

$$\Delta\hat{P}_{\text{O-UE,inc}}^{\mathrm{RMT,n}}(k,\hat{v}_k)=\Delta\hat{P}_{\mathrm{UE,inc}}^{\mathrm{RMT,n}}(k,\hat{v}_k)-\Delta\hat{P}_{\mathrm{OE,dec}}^{\mathrm{RMT,n}}(k,\hat{v}_k).$$
(65)

**Step 3:**

Then, through comparing $\Delta\hat{P}_{\text{O-UE, inc}}^{\mathrm{RMT,s}}(k,\hat{v}_k)$ in (53) with $\Delta\hat{P}_{\text{O-UE,inc}}^{\mathrm{RMT,n}}(k,\hat{v}_k)$ in (65), the proposed RMT-ADC estimator adaptively selects its decision criterion between (18) and (54) and determines whether $\hat{v}_k$ should be incorporated into the **selected** decision criterion in the following way:

(1) If $\Delta\hat{P}_{\text{O-UE,inc}}^{\mathrm{RMT,n}}(k,\hat{v}_k)\ge\Delta\hat{P}_{\text{O-UE, inc}}^{\mathrm{RMT,s}}(k,\hat{v}_k)$, the increased under-estimation probability of the RMT estimator incurred by $\hat{v}_k$ under the assumption that $l_k$ is arising from noise is greater than or equal to its increased under-estimation probability incurred by $\hat{v}_k$ under the assumption that $l_k$ is arising from a signal. Consequently, we can infer that the noise variance should be estimated under the assumption that $l_k$ is arising from noise, and thus the decision criterion (54) should be selected. Moreover, the proposed RMT-ADC estimator should utilize $\Delta\hat{P}_{\text{O-UE,inc}}^{\mathrm{RMT,n}}(k,\hat{v}_k)$ to determine whether $\hat{v}_k$ should be incorporated into the decision criterion (54) as follows:

(a) If $\Delta\hat{P}_{\text{O-UE,inc}}^{\mathrm{RMT,n}}(k,\hat{v}_k)\le 0$, i.e., $\Delta\hat{P}_{\mathrm{OE,dec}}^{\mathrm{RMT,n}}(k,\hat{v}_k)\ge\Delta\hat{P}_{\mathrm{UE,inc}}^{\mathrm{RMT,n}}(k,\hat{v}_k)$, the decreased over-estimation probability of the RMT estimator incurred by $\hat{v}_k$ is greater than or equal to its increased under-estimation probability incurred by $\hat{v}_k$. Therefore, the proposed RMT-ADC estimator should utilize $\Delta\hat{P}_{\mathrm{OE,dec}}^{\mathrm{RMT,n}}(k,\hat{v}_k)$ to determine whether $\hat{v}_k$ should be incorporated into the decision criterion (54) as follows:

① If $\Delta\hat{P}_{\mathrm{OE,dec}}^{\mathrm{RMT,n}}(k,\hat{v}_k)\le 0$, $\hat{v}_k$ should be incorporated into the decision criterion (54), and thus the decision criterion in (55) should be selected.

Then, $H_k$ is accepted if (55) is satisfied, and $k$ is increased by one. Otherwise, the number of signals is estimated as $\hat{q}=k-1$.

② Otherwise, i.e., $\Delta\hat{P}_{\mathrm{OE,dec}}^{\mathrm{RMT,n}}(k,\hat{v}_k)>0$, $\hat{v}_k$ should not be incorporated into the decision criterion (54).

Then, $H_k$ is accepted if (54) is satisfied, and $k$ is increased by one. Otherwise, the number of signals is estimated as $\hat{q}=k-1$.

(b) Otherwise, $\Delta\hat{P}_{\mathrm{OE,dec}}^{\mathrm{RMT,n}}(k,\hat{v}_k)<\Delta\hat{P}_{\mathrm{UE,inc}}^{\mathrm{RMT,n}}(k,\hat{v}_k)$, the increased under-estimation probability of the RMT estimator incurred by $\hat{v}_k$ is greater than its decreased over-estimation probability incurred by $\hat{v}_k$. Therefore, the proposed RMT-ADC estimator should utilize $\Delta\hat{P}_{\mathrm{UE,inc}}^{\mathrm{RMT,n}}(k,\hat{v}_k)$ to determine whether $\hat{v}_k$ should be incorporated into the decision criterion in (54) as follows:

① If $\Delta\hat{P}_{\mathrm{UE,inc}}^{\mathrm{RMT,n}}(k,\hat{v}_k)\ge 0$, $\hat{v}_k$ should be incorporated into the decision criterion (54), and thus the decision criterion (55) should be selected.

Then, $H_k$ is accepted if (55) is satisfied, and $k$ is increased by one. Otherwise, the number of signals is estimated as $\hat{q}=k-1$.

② Otherwise, i.e., $\Delta\hat{P}_{\mathrm{UE,inc}}^{\mathrm{RMT,n}}(k,\hat{v}_k)<0$, $\hat{v}_k$ should not be incorporated into the decision criterion (54).

Then, $H_k$ is accepted if (54) is satisfied, and $k$ is increased by one. Otherwise, the number of signals is estimated as $\hat{q}=k-1$.

(2) Otherwise, $\Delta\hat{P}_{\text{O-UE, inc}}^{\mathrm{RMT,n}}(k,\hat{v}_k)<\Delta\hat{P}_{\text{O-UE, inc}}^{\mathrm{RMT,s}}(k,\hat{v}_k)$, the increased under-estimation probability of the RMT estimator incurred by $\hat{v}_k$ under the assumption that $l_k$ is arising from a signal is greater than its increased under-estimation probability incurred by $\hat{v}_k$ under the assumption that $l_k$ is arising from noise. Consequently, the RMT estimator down-estimates the number of signals. However, the b-RMT estimator may still overestimate the number of signals due to the bias term $\hat{v}_k$. Therefore, in order to overcome both the higher under-estimation probability of the RMT estimator and the higher over-estimation probability of the b-RMT estimator, we should proceed to Step 4 to further test whether $l_k$ is arising from a signal or from noise.

**Step 4:**

According to (49) and (52), the overall decreased over-estimation probability of the RMT estimator incurred by $\hat{v}_k$ under the assumption that $l_k$ is arising from a signal is derived as

$$\Delta\hat{P}^{\text{RMT,s}}_{\text{O-OE, dec}}(k,\hat{v}_k)=\Delta\hat{P}^{\text{RMT,s}}_{\text{OE, dec}}(k,\hat{v}_k)-\Delta\hat{P}^{\text{RMT,s}}_{\text{UE,inc}}(k,\hat{v}_k) \quad . \tag{66}$$

Then, through comparing $\Delta\hat{P}^{\text{RMT,s}}_{\text{O-OE, dec}}(k,\hat{v}_k)$ in (66) with $\Delta\hat{P}^{\text{RMT,n}}_{\text{O-UE,inc}}(k,\hat{v}_k)$ in (65), the proposed RMT-ADC estimator adaptively selects its decision criterion between (18) and (54) and determines whether $\hat{v}_k$ should be incorporated into the selected decision criterion in the following way:

(1) If $\Delta\hat{P}^{\text{RMT,n}}_{\text{O-UE,inc}}(k,\hat{v}_k)\geq\Delta\hat{P}^{\text{RMT,s}}_{\text{O-OE, dec}}(k,\hat{v}_k)$, the increased under-estimation probability of the RMT estimator incurred by $\hat{v}_k$ when assuming that $l_k$ is arising from noise is greater than or equal to its decreased over-estimation probability incurred by $\hat{v}_k$ when assuming that $l_k$ is arising a signal. Therefore, the noise variance should be estimated under the assumption that $l_k$ is arising from noise, and thus the decision criterion (54) should be selected. Moreover, the proposed RMT-ADC estimator should utilize $\Delta\hat{P}^{\text{RMT,n}}_{\text{O-UE,inc}}(k,\hat{v}_k)$ to adaptively determine whether $\hat{v}_k$ should be incorporated into the selected decision criterion (54) as follows:

(a) If $\Delta\hat{P}^{\text{RMT,n}}_{\text{O-UE,inc}}(k,\hat{v}_k)\leq 0$, i.e., $\Delta\hat{P}^{\text{RMT,n}}_{\text{OE,dec}}(k,\hat{v}_k)\geq\Delta\hat{P}^{\text{RMT,n}}_{\text{UE,inc}}(k,\hat{v}_k)$, the decreased over-estimation probability of the RMT estimator incurred by $\hat{v}_k$ is greater than or equal to its increased under-estimation probability incurred by $\hat{v}_k$ under the assumption that $l_k$ is arising from noise. Therefore, the proposed RMT-ADC estimator should utilize $\Delta\hat{P}^{\text{RMT,n}}_{\text{OE,dec}}(k,\hat{v}_k)$ to determine whether $\hat{v}_k$ should be incorporated into the selected decision criterion (54) as follows:

① If $\Delta\hat{P}^{\text{RMT,n}}_{\text{OE,dec}}(k,\hat{v}_k)\leq 0$, $\hat{v}_k$ should be incorporated into the decision criterion (54), and thus the decision criterion in (55) should be selected.

Then, $H_k$ is accepted if (55) is satisfied, and $k$ is increased by one. Otherwise, the number of signals is estimated as $\hat{q}=k-1$.

② Otherwise, $\Delta\hat{P}^{\text{RMT,n}}_{\text{OE,dec}}(k,\hat{v}_k)>0$, $\hat{v}_k$ should not be incorporated into the decision criterion (54).

Then, $H_k$ is accepted if (54) is satisfied, and $k$ is increased by one. Otherwise, the number of signals is estimated as $\hat{q}=k-1$.

(b) Otherwise, $\Delta\hat{P}^{\text{RMT,n}}_{\text{OE,dec}}(k,\hat{v}_k)<\Delta\hat{P}^{\text{RMT,n}}_{\text{UE,inc}}(k,\hat{v}_k)$, the proposed RMT-ADC estimator should utilize $\Delta\hat{P}^{\text{RMT,n}}_{\text{UE,inc}}(k,\hat{v}_k)$ to determine whether $\hat{v}_k$ should be incorporated into the decision criterion (54) as follows:

① If $\Delta\hat{P}^{\text{RMT,n}}_{\text{UE,inc}}(k,\hat{v}_k)\geq 0$, $\hat{v}_k$ should be incorporated into the decision criterion in (54), and thus the decision criterion in (55) should be selected.

Then, $H_k$ is accepted if (55) is satisfied, and $k$ is increased by one. Otherwise, the number of signals is estimated as $\hat{q}=k-1$.

② Otherwise, $\Delta\hat{P}^{\text{RMT,n}}_{\text{UE,inc}}(k,\hat{v}_k)<0$, $\hat{v}_k$ should not be incorporated into the decision criterion in (54).

Then, $H_k$ is accepted if (54) is satisfied, and $k$ is increased by one. Otherwise, the number of signals is estimated as $\hat{q}=k-1$.

(2) Otherwise, $\Delta\hat{P}^{\text{RMT,n}}_{\text{O-UE,inc}}(k,\hat{v}_k)<\Delta\hat{P}^{\text{RMT,s}}_{\text{O-OE, dec}}(k,\hat{v}_k)$, the decreased over-estimation probability of the RMT estimator incurred by $\hat{v}_k$ when assuming that $l_k$ is arising from a signal is greater than its increased under-estimation probability incurred by $\hat{v}_k$ when assuming that $l_k$ is arising from noise. Therefore, the noise variance should be estimated under the assumption that $l_k$ is arising from a signal, and thus the decision criterion (18) should be selected. Moreover, the proposed RMT-ADC estimator should utilize $\Delta\hat{P}^{\text{RMT,s}}_{\text{O-OE, dec}}(k,\hat{v}_k)$ to adaptively determine whether $\hat{v}_k$ should be incorporated into the selected decision criterion (18) as follows:

(a) If $\Delta\hat{P}^{\text{RMT,s}}_{\text{O-OE, dec}}(k,\hat{v}_k)\geq 0$, i.e., $\Delta\hat{P}^{\text{RMT,s}}_{\text{OE, dec}}(k,\hat{v}_k)\geq\Delta\hat{P}^{\text{RMT,s}}_{\text{UE,inc}}(k,\hat{v}_k)$, the decreased over-estimation probability of the RMT estimator incurred by $\hat{v}_k$ is greater than or equal to its increased under-estimation probability incurred by $\hat{v}_k$ under the assumption that $l_k$ is arising from a signal. Therefore, the proposed RMT-ADC estimator should utilize $\Delta\hat{P}^{\text{RMT,s}}_{\text{OE, dec}}(k,\hat{v}_k)$ to determine whether $\hat{v}_k$ should be incorporated into the decision criterion (18) as follows:

① If $\Delta\hat{P}^{\text{RMT,s}}_{\text{OE, dec}}(k,\hat{v}_k)\leq 0$, $\hat{v}_k$ should be incorporated into the decision criterion (18), and thus the decision criterion (43) should be selected.

Then, $H_k$ is accepted if (43) is satisfied, and $k$ is increased by one. Otherwise, the number of signals is estimated as $\hat{q}=k-1$.

② Otherwise, $\Delta\hat{P}^{\text{RMT,s}}_{\text{OE, dec}}(k,\hat{v}_k)>0$, $\hat{v}_k$ should not be

incorporated into the decision criterion (18).

Then, $H_k$ is accepted if (18) is satisfied, and $k$ is increased by one. Otherwise, the number of signals is estimated as $\hat{q} = k - 1$.

(b) Otherwise, $\Delta\hat{P}^{\text{RMT,s}}_{\text{OE, dec}}(k,\hat{v}_k) < \Delta\hat{P}^{\text{RMT,s}}_{\text{UE,inc}}(k,\hat{v}_k)$, the proposed RMT-ADC estimator should utilize $\Delta\hat{P}^{\text{RMT,s}}_{\text{UE,inc}}(k,\hat{v}_k)$ to determine whether $\hat{v}_k$ should be incorporated into the decision criterion (18) as follows:

① If $\Delta\hat{P}^{\text{RMT,s}}_{\text{UE,inc}}(k,\hat{v}_k) \geq 0$, $\hat{v}_k$ should be incorporated into the decision criterion (18), and thus the decision criterion (43) should be selected.

Then, $H_k$ is accepted if (43) is satisfied, and $k$ is increased by one. Otherwise, the number of signals is estimated as $\hat{q} = k - 1$.

② Otherwise, $\Delta\hat{P}^{\text{RMT,s}}_{\text{UE,inc}}(k,\hat{v}_k) < 0$, $\hat{v}_k$ should not be incorporated into the decision criterion (18).

Then, $H_k$ is accepted if (18) is satisfied, and $k$ is increased by one. Otherwise, the number of signals is estimated as $\hat{q} = k - 1$.

To summarize, the proposed RMT-ADC estimator can adaptively determine whether the noise variance should be estimated under the assumption that the eigenvalue being tested is arising from a signal or noise. Moreover, it can adaptively select (18) or (54) as its decision criterion, and can adaptively determine whether the bias term $\hat{v}_k$ should be incorporated into the selected decision criterion or not. Therefore, the RMT-ADC estimator can successfully detect the signal eigenvalues immersed in the bias term $\hat{v}_k$, and thus overcomes the higher under-estimation probability of the RMT estimator. In addition, the proposed RMT-ADC estimator can also overcome the higher over-estimation probability of the RMT estimator resulted from the

Simulation results are presented to illustrate the performance of the proposed RMT-ADC estimator and compare it with the RMT estimator. In the simulations, we utilize a population covariance matrix $\mathbf{\Sigma} = E[\mathbf{x}\mathbf{x}^H]$ that has $q$ signal components with true signal strength $\lambda = [\lambda_1, \lambda_2, \cdots, \lambda_q]$ and $p - q$ "noise" eigenvalues $\lambda_{q+1} = \cdots = \lambda_p = \sigma^2 = 1$, and $\gamma = p/n$. The performance measure is the probability of over-estimation $P_f = \Pr(\hat{q} > q)$, and all results are averaged over 10,000 independent Monte Carlo runs. Fig.1 shows simulation results for the over-estimation probability $\Pr(\hat{q} > q)$ as a function of number of sensors $p$ with pre-fixed value $\alpha = 0.005$ and $p/n = 1/2$ for an *a priori* known noise variance $\sigma^2$: (a) $q = 0$ and $\lambda = []$; (b) $q = 3$ and $\lambda = [100, 80, 60]$, and Fig. 2 shows the corresponding results for the estimated noise variance $\hat{\sigma}^2$. As can be seen form Fig. 1, the over-estimation probability of the RMT is greater than the pre-fixed value $\alpha = 0.005$ for the case that there is no signal with $\lambda = []$, which is because $\varphi_{n,p-1} < \varphi_{n,p}$. In addition, the over-estimation probability of the RMT is less than the pre-fixed value $\alpha = 0.005$ for the case that there are three signals with $\lambda = [150, 120, 100]$, which is because the bias term $\hat{v}_3$ decreased the over-estimation probability of the RMT estimator. On the contrary, the over-estimation probability of the proposed RMT-ADC estimator remains around the pre-fixed value $\alpha = 0.005$ for both cases for the known noise variance $\sigma^2$. Moreover, as can be seen from Fig. 2, the over-estimation probability of the RMT becomes far greater than the pre-fixed value $\alpha = 0.005$ for the estimated noise level $\hat{\sigma}^2$. This is because the noise variance $\hat{\sigma}^2$ is estimated under the assumption that the eigenvalue being tested is arising from a signal, and thus will be under-estimated when the eigenvalue being tested is actually arising from noise. Then, the under-estimated noise variance $\hat{\sigma}^2$ increases the over-estimation probability of the RMT estimator. On the contrary, the over-estimation probability of the proposed RMT-ADC estimator remains around the pre-fixed value $\alpha = 0.005$ for both cases for the estimated noise variance $\hat{\sigma}^2$.

As can be seen from these results, the proposed RMT-ADC estimator significantly outperforms the RMT estimator.

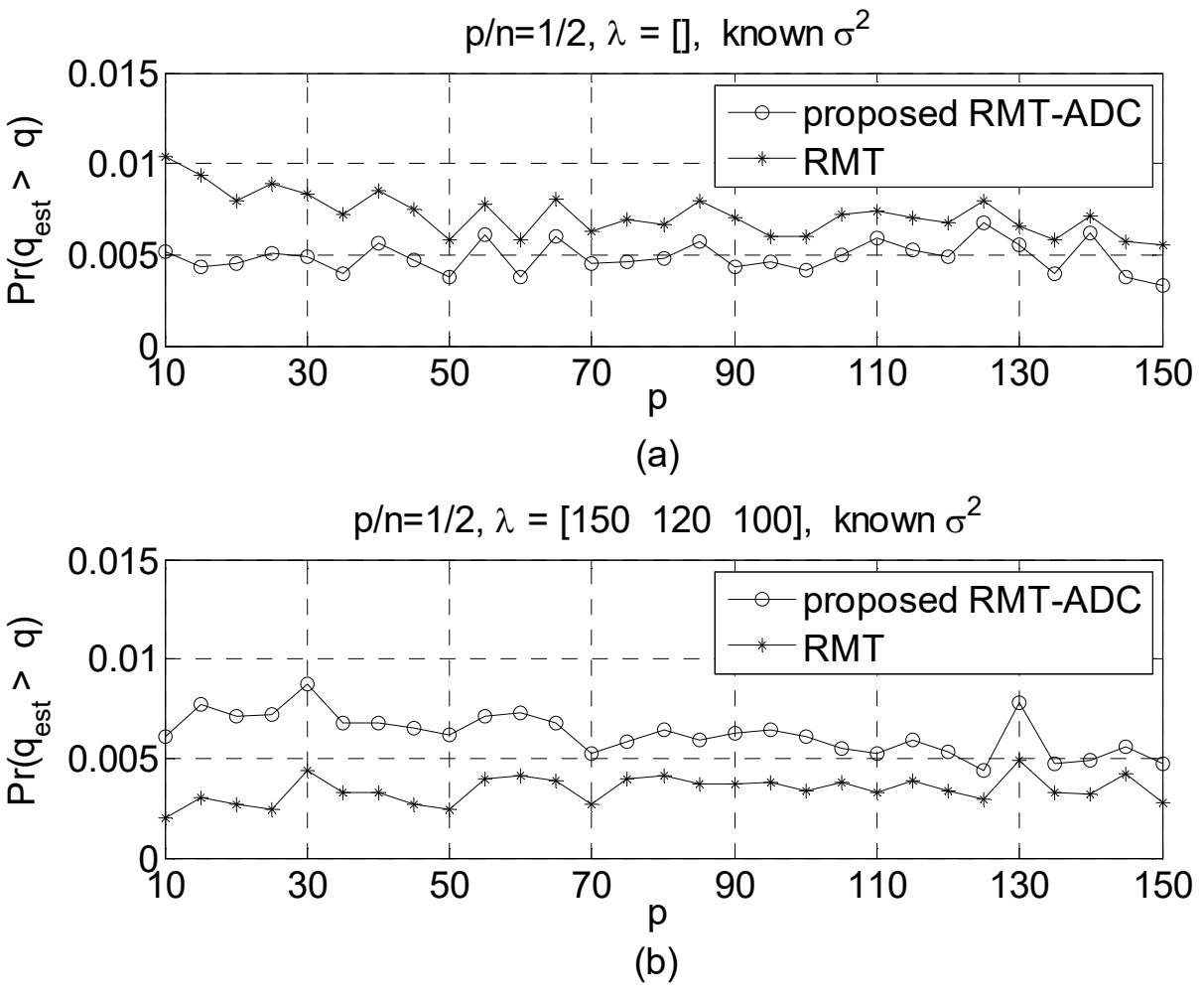

Fig.1. Over-estimation probability $\Pr(\hat{q} > q)$ as a function of number of sensors $p$ with pre-fixed value $\alpha = 0.005$ and $p/n = 1/2$ for an *a priori* known noise variance $\sigma^2$: (a) $q = 0$ and $\lambda = []$, (b) $q = 3$ and $\lambda = [150, 120, 100]$.

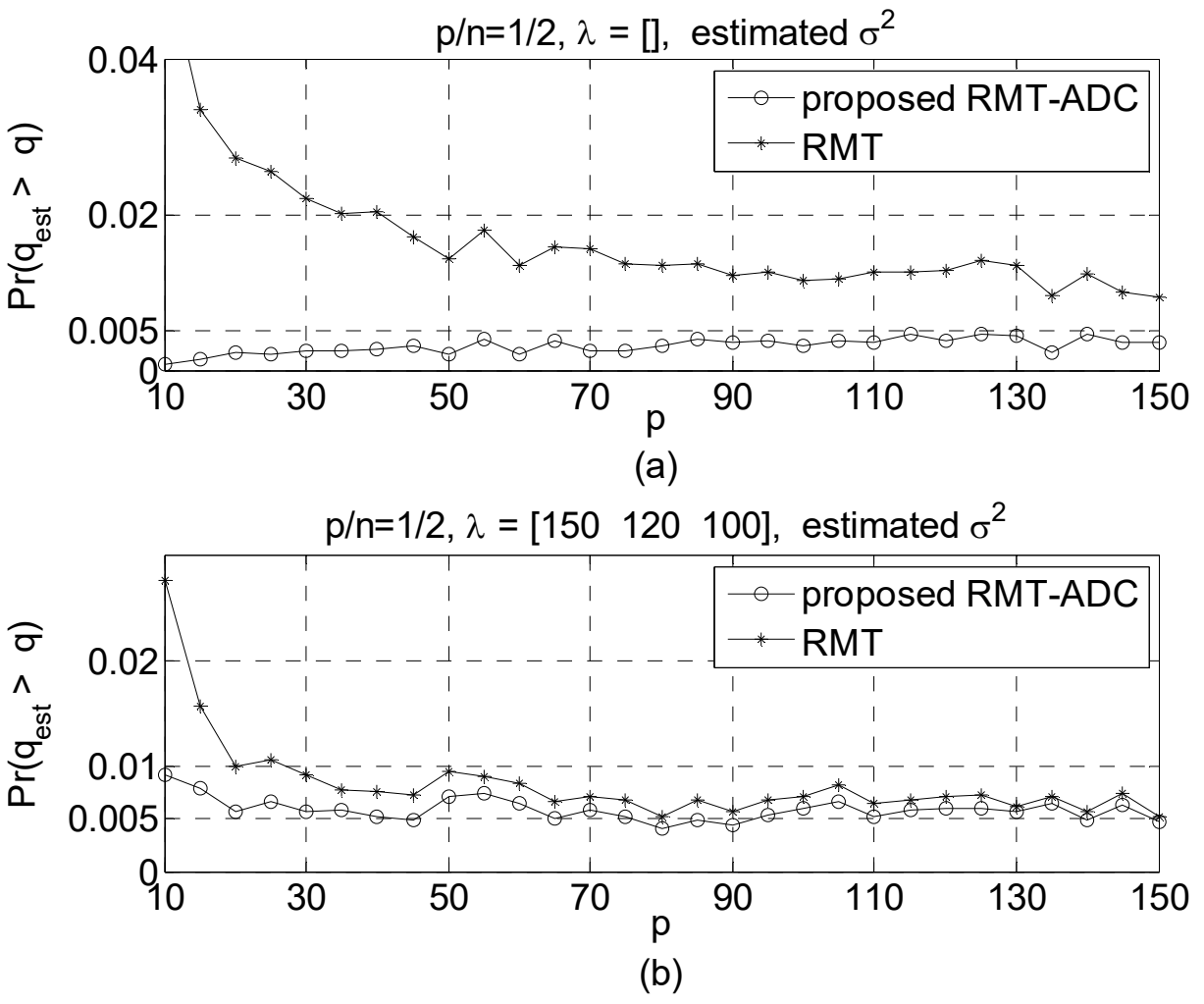

Fig.2. Over-estimation probability $\Pr(\hat{q} > q)$ as a function of number of sensors $p$ with pre-fixed value $\alpha = 0.005$ and $p / n = 1 / 2$ for estimated noise variance $\hat{\sigma}^2$ : (a) $q = 0$ and $\lambda = []$, (b) $q = 3$ and $\lambda = [150, 120, 100]$.

## IV. Simulation and Discussions

In this section, we examine the performance of the proposed RMT-ADC estimator in Section III.B, and compare it with the standard MDL and AIC estimators [7]-[8], modified AIC estimator [25] with $C = 2$, and the RMT estimator in (18) using Monte Carlo simulations. For all simulations, we assume real valued signals and real valued Gaussian noise, the significant level $\alpha$ in both the RMT estimator in (18) and the proposed RMT-ADC estimator is set as $\alpha = 0.005$, and we use a population covariance matrix $\mathbf{\Sigma} = E[\mathbf{x}\mathbf{x}^H]$ that has $q$ unknown signal components with true signal strength $\lambda = [\lambda_1, \lambda_2, \cdots, \lambda_q]$ and $p - q$ "noise" eigenvalues $\lambda_{q+1} = \cdots = \lambda_p = \sigma^2 = 1$. The performance measures are the probability of misdetection $\Pr(\hat{q} \neq q)$, the probability of under-estimation $\Pr(\hat{q} < q)$, and the probability of the over-estimation $\Pr(\hat{q} > q)$. All results are averaged over 12,000 independent Monte Carlo runs.

### *A. Performance of the Proposed RMT-ADC Estimator for the case $p / n < 1$*

Firstly, we examine the over-estimation probability of the proposed RMT-ADC estimator for the case $p / n < 1$. In this simulation, we consider two cases: (a) the case without signals; and (b) the case with three strong signals. In these cases, the misdetection probability is mainly the over-estimation probability.

Fig. 3 shows the over-estimation probability of the proposed RMT-ADC estimator as a function of the system size $p$ with fixed ratio $p / n = 1 / 2$ for the case when there is no signal with $\lambda = []$, and Fig. 4 shows the corresponding results for the case when there are $q{=}3$ strong signals with $\lambda = [150, 120, 100]$. For comparison, the results for the standard MDL and AIC estimators, modified AIC estimator, and the RMT estimator [24] are also shown in these Figures.

As can be seen from Fig. 3 and Fig. 4, the over-detection probability of the proposed RMT-ADC estimator is about the pre-fixed value $\alpha = 0.005$ for both cases, while the over-estimation probability of RMT estimator is much higher than the pre-fixed value $\alpha = 0.005$ especially when the system size $p$ is relatively small. This is because the noise variance $\hat{\sigma}^2$ is estimated under the assumption that the eigenvalue being tested is arising from a signal, and thus will be under-estimated when the eigenvalue being tested is actually arising from noise. Then, the under-estimated noise variance $\hat{\sigma}^2$ increases the over-estimation probability of the RMT estimator. Therefore, the RMT-ADC estimator overcomes the higher over-estimation probability of the RMT estimator. Moreover, though the over-estimation probability of the MDL estimator is about 0.005 when the system size $p = 10$ for the case when there is no signal with $\lambda = []$ (as shown in Fig. 3), it will becomes 0.06 for the case when there are $q{=}3$ strong signals with $\lambda = [150, 120, 100]$ (as shown in Fig. 4). In addition, though the modified AIC estimator has zero over-estimation probability for these two cases, but Fig. 5 will show that it has much larger under-estimation probability when the system size $p$ is relatively small as signal eigenvalues will be immersed in the bias term among eigenvalues. Finally, the conventional AIC estimator has non-negligible over-estimation probability, especially when the system size $p$ is relatively small.

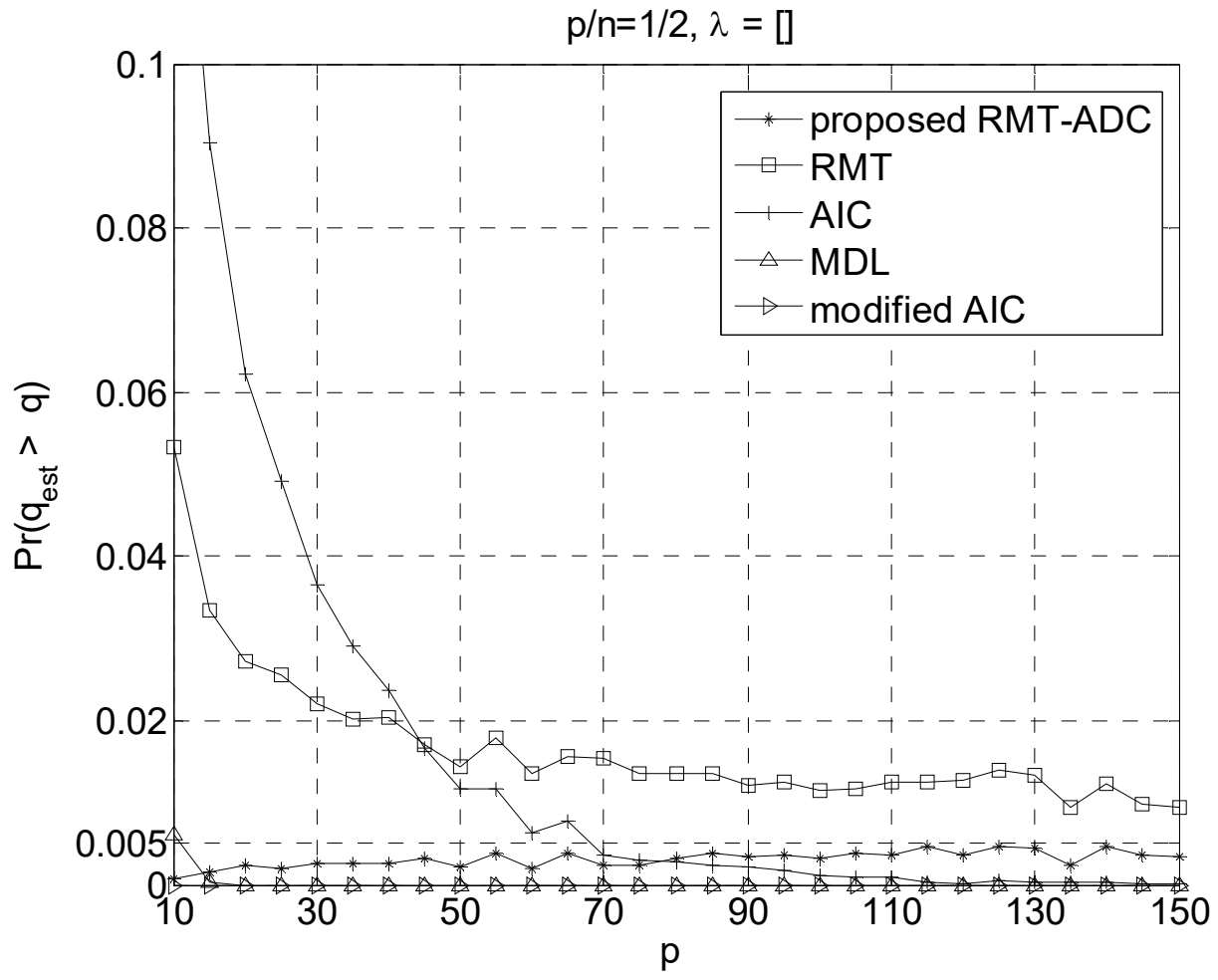

Fig. 3. Comparison of over-estimation probability of various algorithms as a function of system size $p$ with fixed ratio

$p / n = 1 / 2$ for the case when there is no signal with $\lambda = []$.

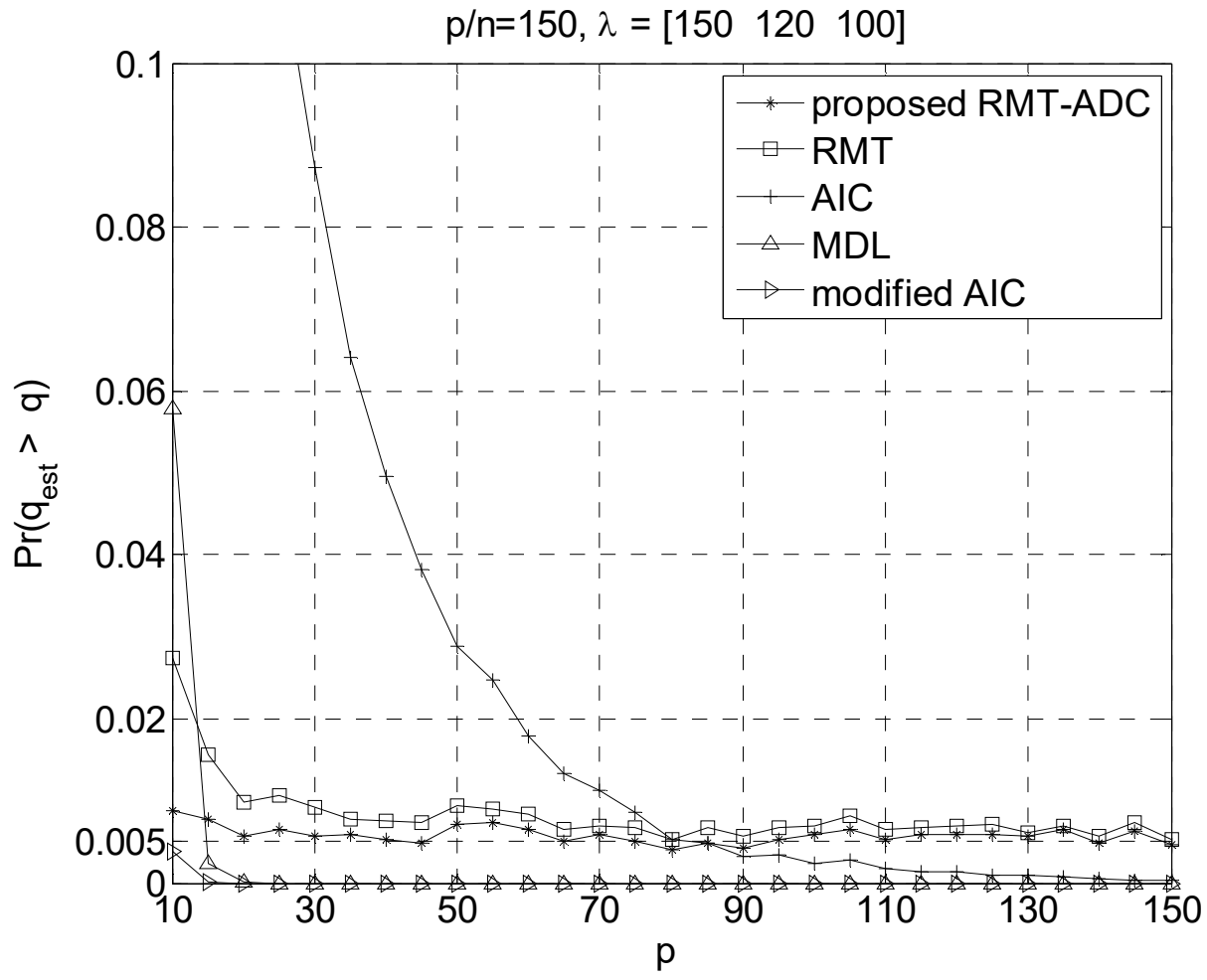

Fig. 4. Comparison of over-estimation probability of various algorithms as a function of system size $p$ with fixed ratio $p / n = 1 / 2$ for the case when there are $q$=3 strong signals with $\lambda = [150, 120, 100]$.

Secondly, we illustrate the detection performance of the proposed RMT-ADC estimator for the case when there are multiple signals. Fig. 5 (a) and (b), respectively, show the mis-estimation probability and over-estimation probability of various algorithms as function of the system size $p$ with fixed ratio $p / n = 1 / 2$ for the case when there are eleven signals with $\lambda = [12,10,10,9, 7, 7, 6, 6, 5, 5,4]$. As can be seen from Fig. 5, the proposed RMT-ADC estimator has much better detection performance than the RMT estimator for small to moderate system size. This is because the RMT-ADC estimator can successfully detect the signal eigenvalue immersed in the bias term among eigenvalues, while the RMT estimator can't. Moreover, the MDL estimator and the modified AIC estimator have much larger under-detection probability than the proposed RMT-ADC estimator. In addition, though the AIC estimator has lower mis-detection probability than the RMT-ADC estimator, it has much larger non-negligible over-estimation probability than the RMT-ADC estimator, especially when the system size $p$ is relatively small, as shown in Fig. 5(b).

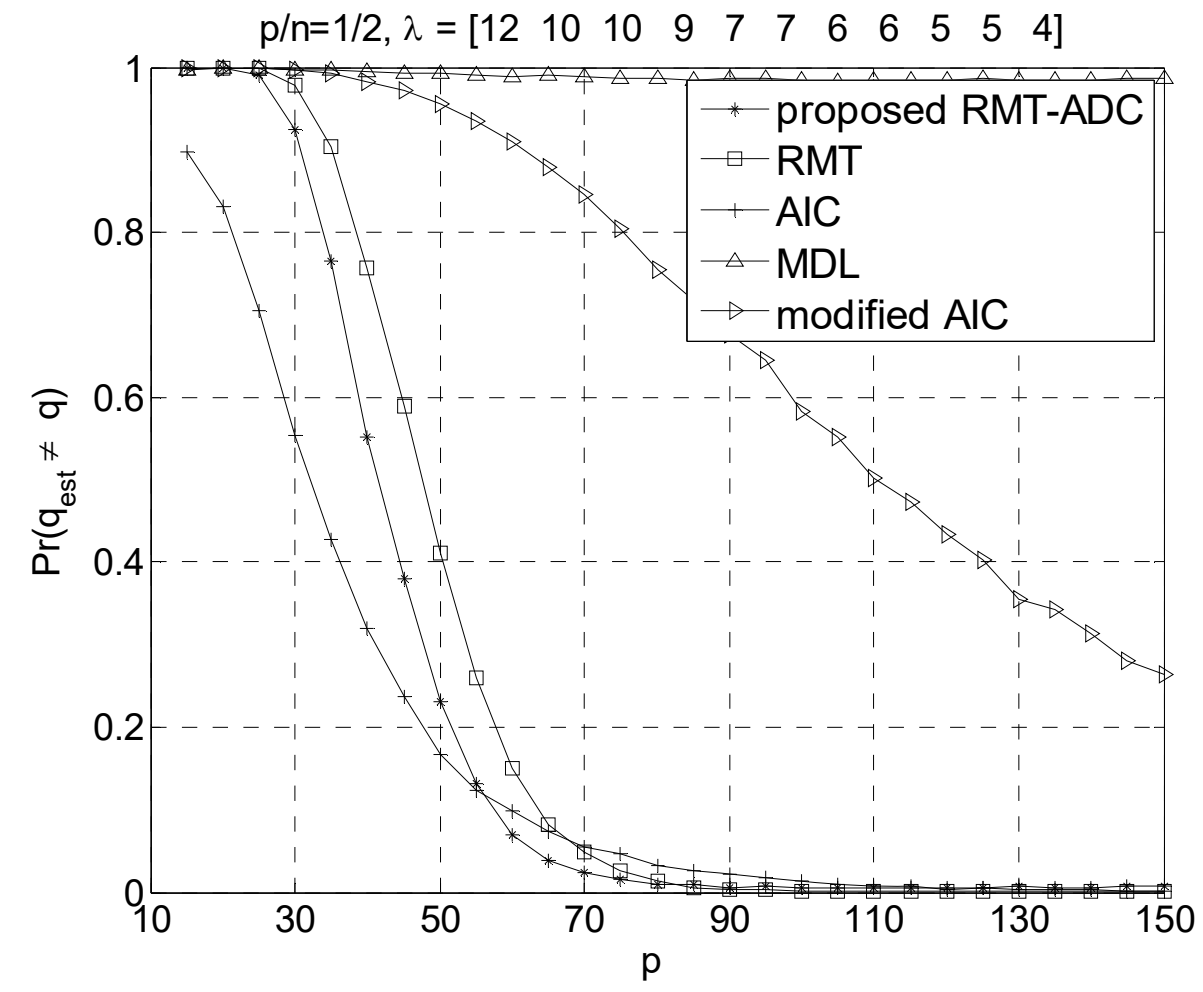

(a) mis-estimation probability

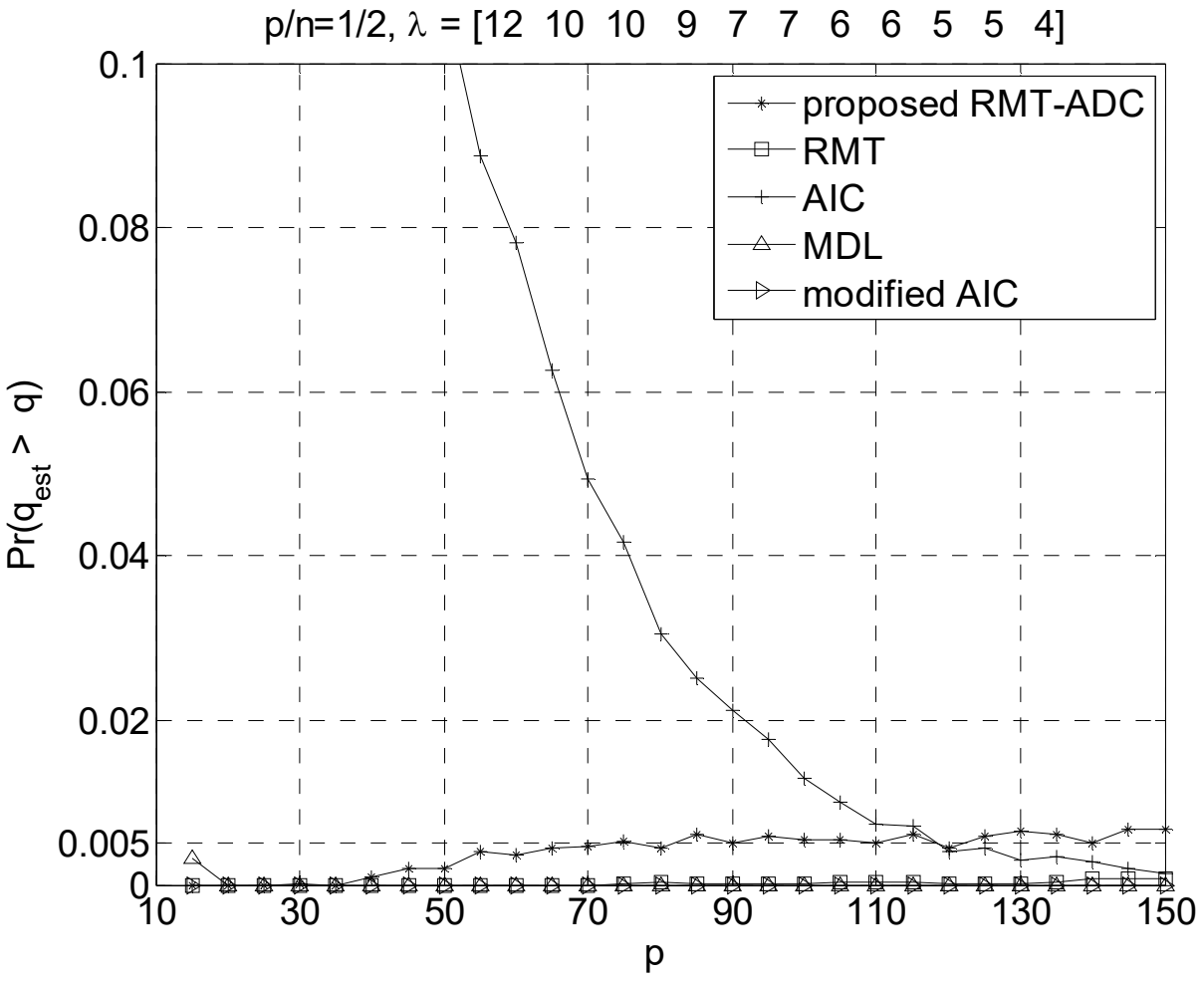

(b) over-estimation probability

Fig. 5. Comparison of (a) mis-estimation probability and (b) over-estimation probability of various algorithms as a function of the system size $p$ with fixed ratio $p / n = 1 / 2$ for the case when there are eleven signals with $\lambda = [12,10,10,9, 7, 7, 6, 6, 5, 5,4]$.

### B. *Performance of the proposed RMT-ADC estimator for the case $p/n > 1$*

Firstly, we examine the over-estimation probability of the proposed RMT-ADC estimator for the case $p / n > 1$. In this simulation, we consider two cases: (a) the case without signals; and (b) the case with one strong signal. In these cases, the misdetection probability mainly comes from the over-estimation probability.

Fig. 6 shows the comparison of over-detection probability of various algorithms as a function of system size $p$ with fixed ratio $p / n = 2$ for the case when there is no signal with $\lambda = []$, and Fig. 7 shows the comparison of (a) misdetection probability and (b) over-detection probability of various algorithms as a function of the system size $p$ with fixed ratio $p/n = 2$ when there are three strong signals with $\lambda = [150, 120, 100]$. As can be seen from Fig. 6 and Fig. 7, the over-detection probability of the proposed RMT-ADC estimator is around the pre-fixed value $\alpha = 0.005$, while the over-estimation probability of RMT estimator is much higher than the pre-fixed value $\alpha = 0.005$, especially when the system size $p$ is relatively small. Therefore, the RMT-ADC estimator overcomes the higher over-estimation probability of the RMT estimator. In addition, though the MDL estimator, AIC estimator and the modified AIC estimator has zero over-estimation probability, the Fig. 7 shows that their under-estimation probability is 100% when there are three strong signals.

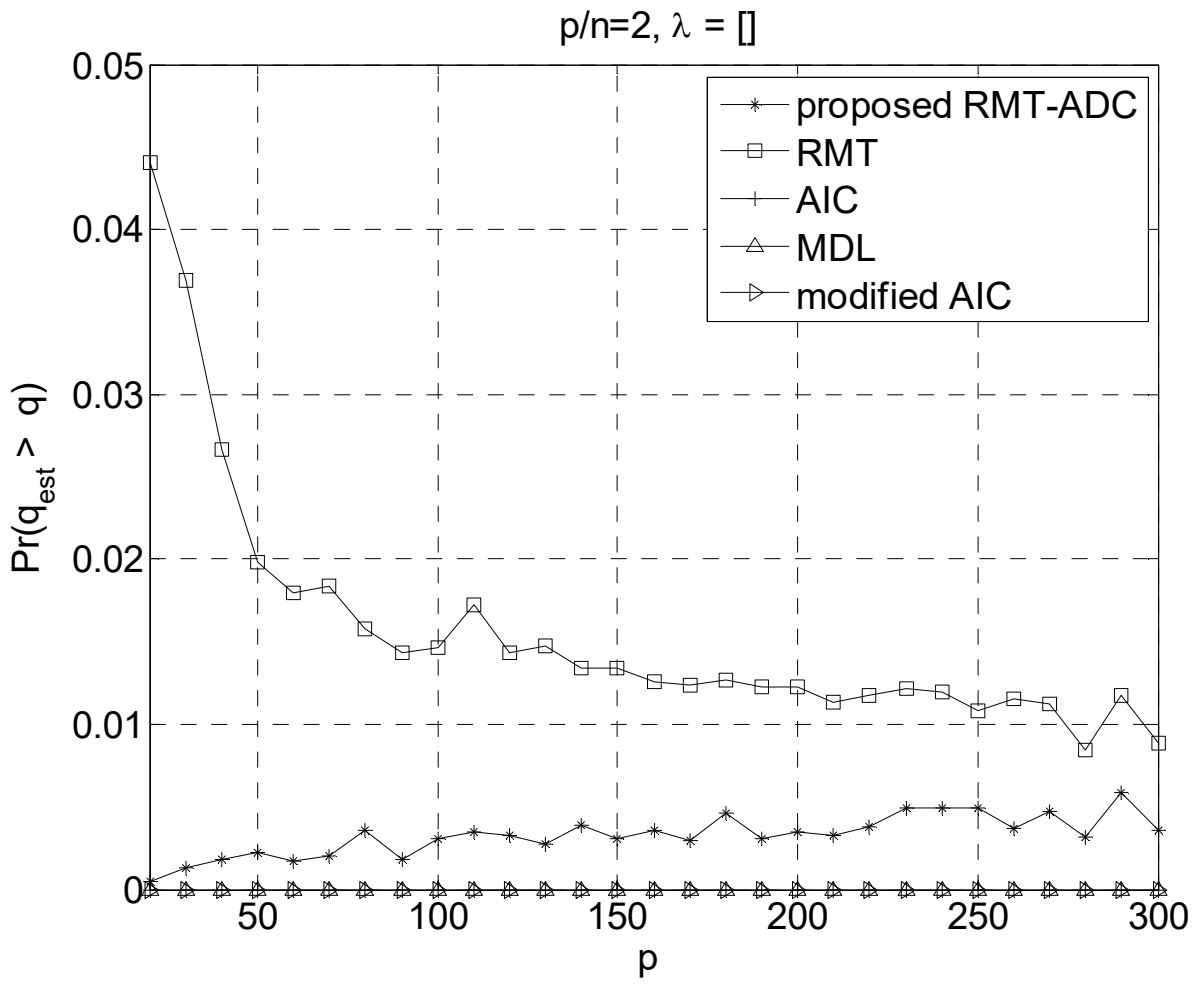

Fig. 6. Comparison of over-detection probability of various algorithms as a function of system size $p$ with fixed ratio $p / n = 2$ when there is no signal with $\lambda = []$.

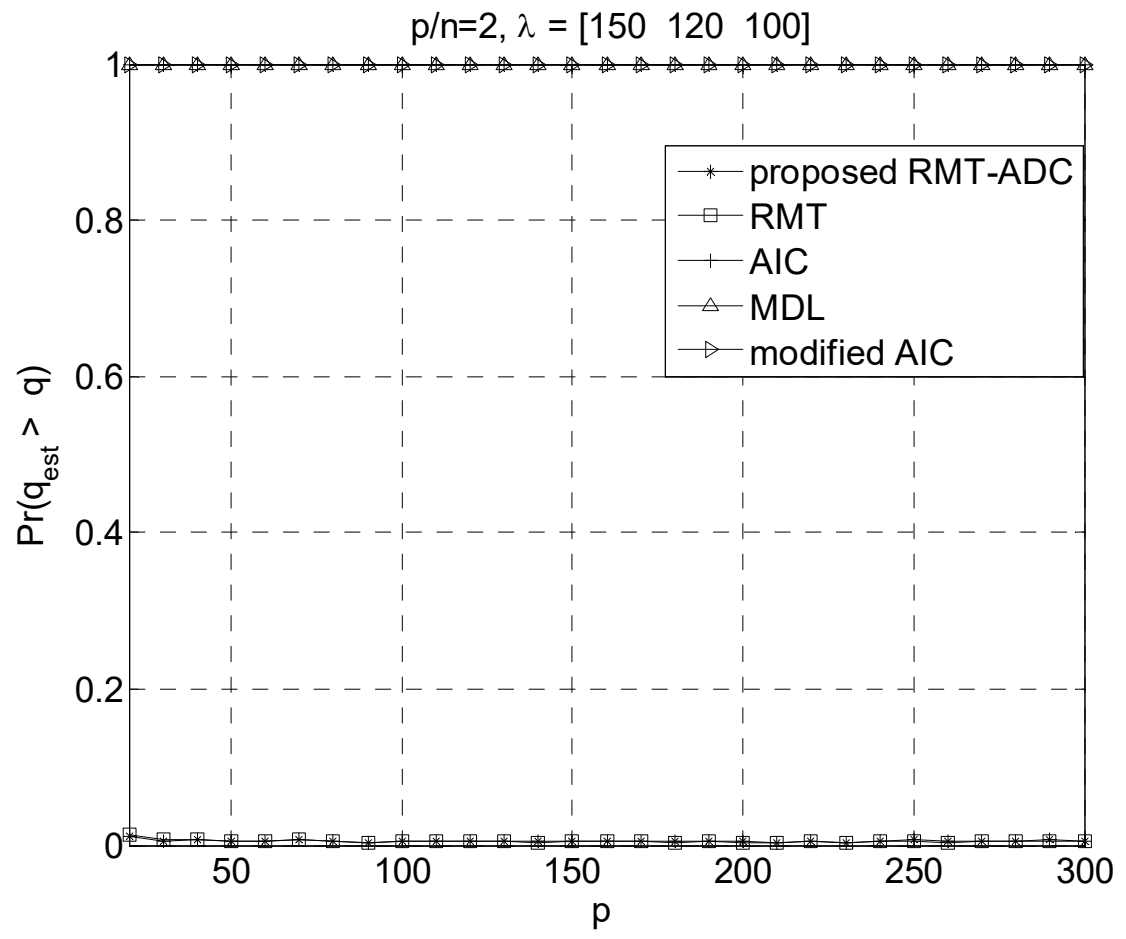

(a) Misdetection probability

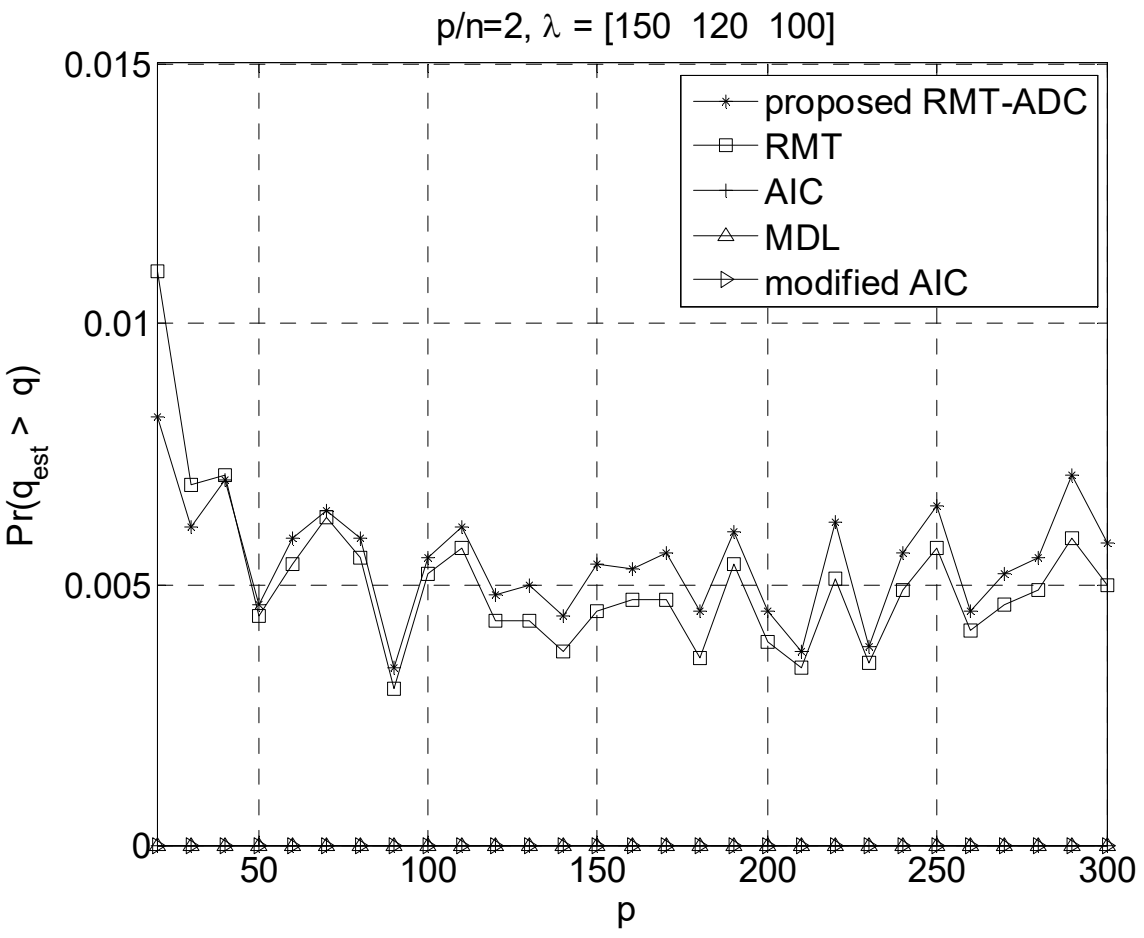

(b) Over-detection probability

Fig. 7. Comparison of (a) mis-estimation probability and (b) over-estimation probability of various algorithms as a function of the system size $p$ with fixed ratio $p / n = 2$ when there is one strong signal with $\lambda = [150, 120, 100]$.

Secondly, we illustrate the detection performance of the proposed RMT-ADC estimator for the case when there are multiple signals. Fig. 8 (a) and (b), respectively, show the misdetection probability and over-estimation probability of various algorithms as function of the system size $p$ with fixed ratio $p/n = 2$ for the case when there are nine strong signals with $\lambda = [16, 15, 12, 12, 12, 10, 10, 8, 6]$.

As can be seen from Fig. 8, the proposed RMT-ADC estimator has much better detection performance (with an improvement up to 24%) than the RMT estimator for relatively small system size, and its over-estimation probability is around the pre-fixed value $\alpha = 0.005$ as the system size $p$ becomes large. In addition, the under-estimation probability of the MDL estimator, the AIC estimator and the modified AIC estimator is 100% in this case.

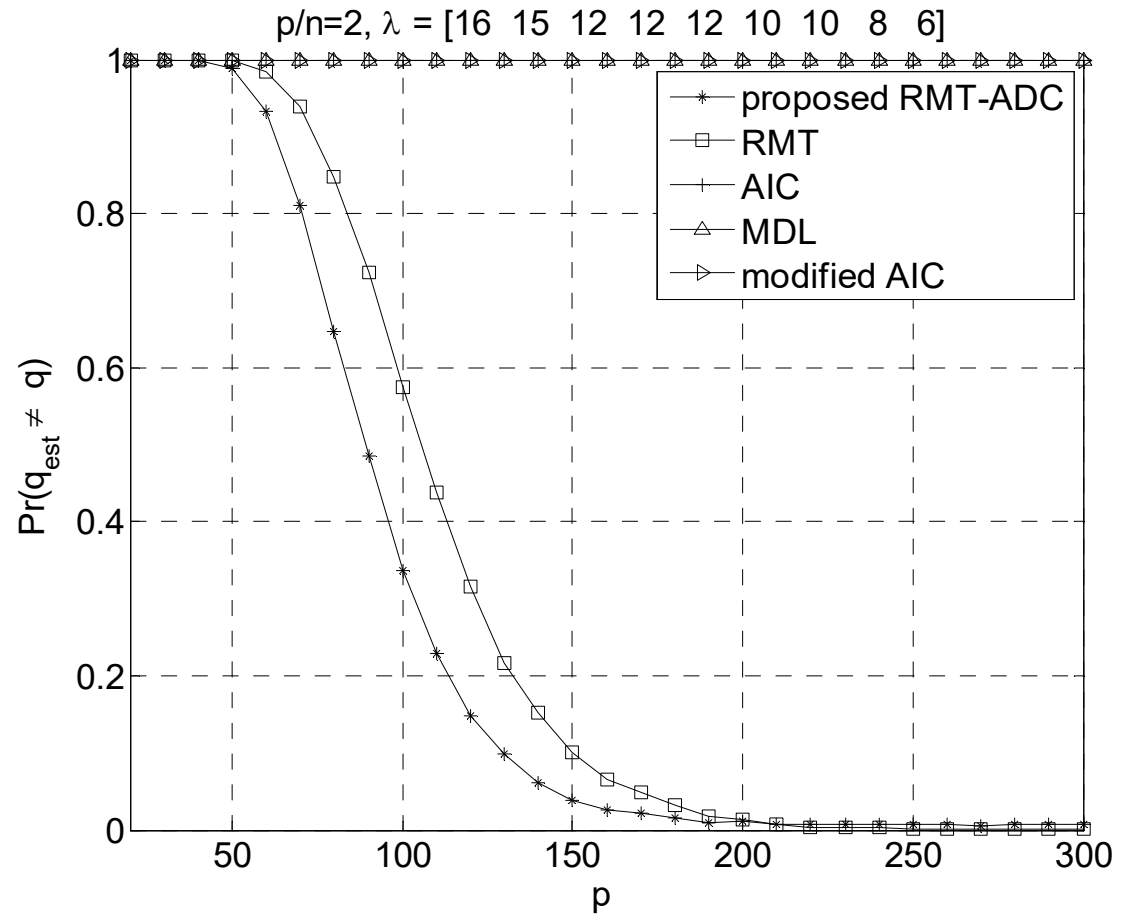

(a)

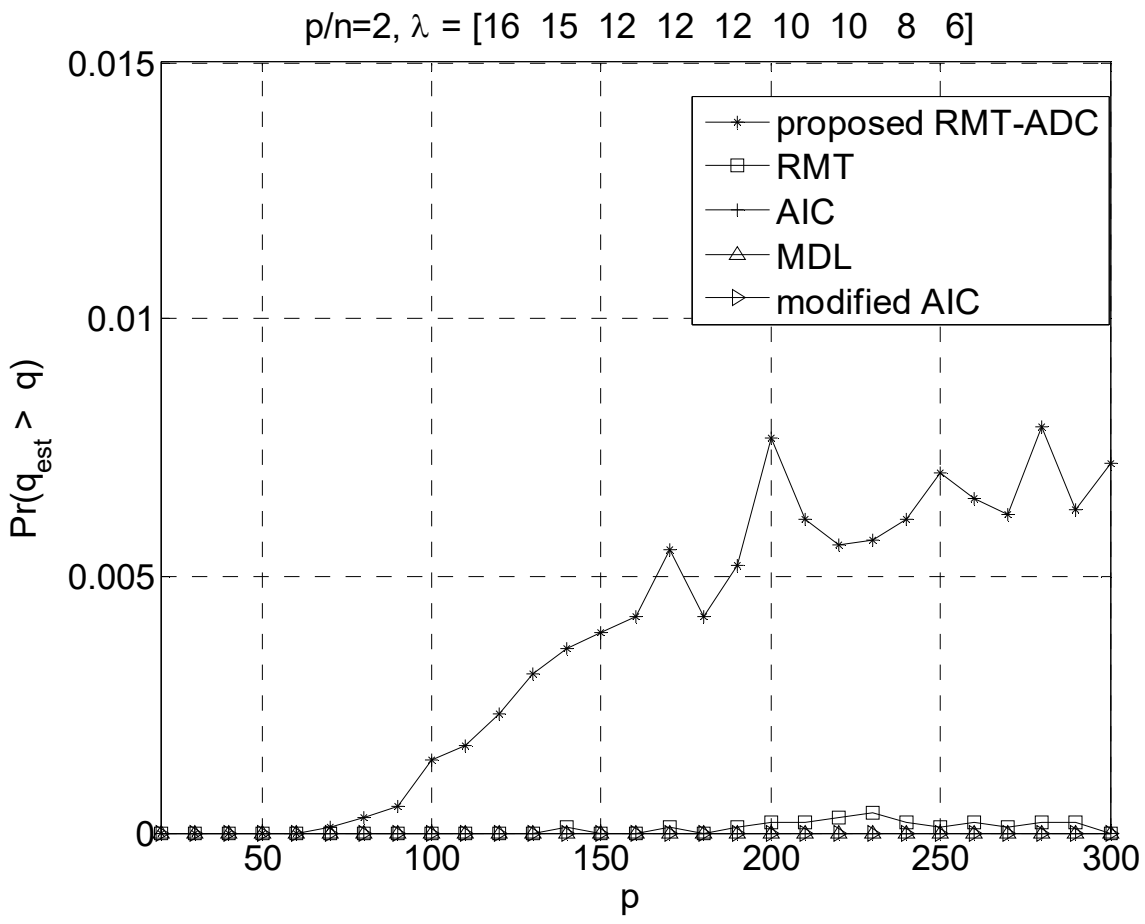

(b)

Fig. 8. Comparison of (a) mis-estimation probability and (b) over-estimation probability of various algorithms as a function of the system size $p$ with fixed ratio $p/n = 2$ for the case when there are nine strong signals with $\lambda = [16, 15, 12, 12, 12, 10, 10, 8, 6]$.

### C. *Performance of the proposed RMT-ADC estimator for various sample size when the system size is fixed*

Firstly, we examine the over-estimation probability of the proposed RMT-ADC estimator when the system size $p$ is fixed. Fig. 9 shows the simulation results for the misdetection probability as a function of sample size $n$ when $p = 50$ : (a) no signal with $\lambda = []$ ; (b) three strong signals with $\lambda = [100, 80, 60]$ . As can be seen from Fig. 9, the over-estimation probability of the proposed RMT-ADC estimator is around the pre-fixed value $\alpha = 0.005$, while the over-estimation probability of the RMT estimator is much higher than the pre-fixed value $\alpha = 0.005$. This is to say, the RMT-ADC estimator overcomes the higher over-estimation probability of the RMT estimator. In addition, the conventional AIC estimator has non-negligible over-estimation probability in these cases. Moreover, though the MDL estimator and the modified AIC estimator has zero over-estimation probability in these cases, but they will have much larger under-estimation probability than the RMT-ADC estimator when there are weak signals immersed in the bias term among eigenvalues, which will be shown in following simulations as in Fig. 10.

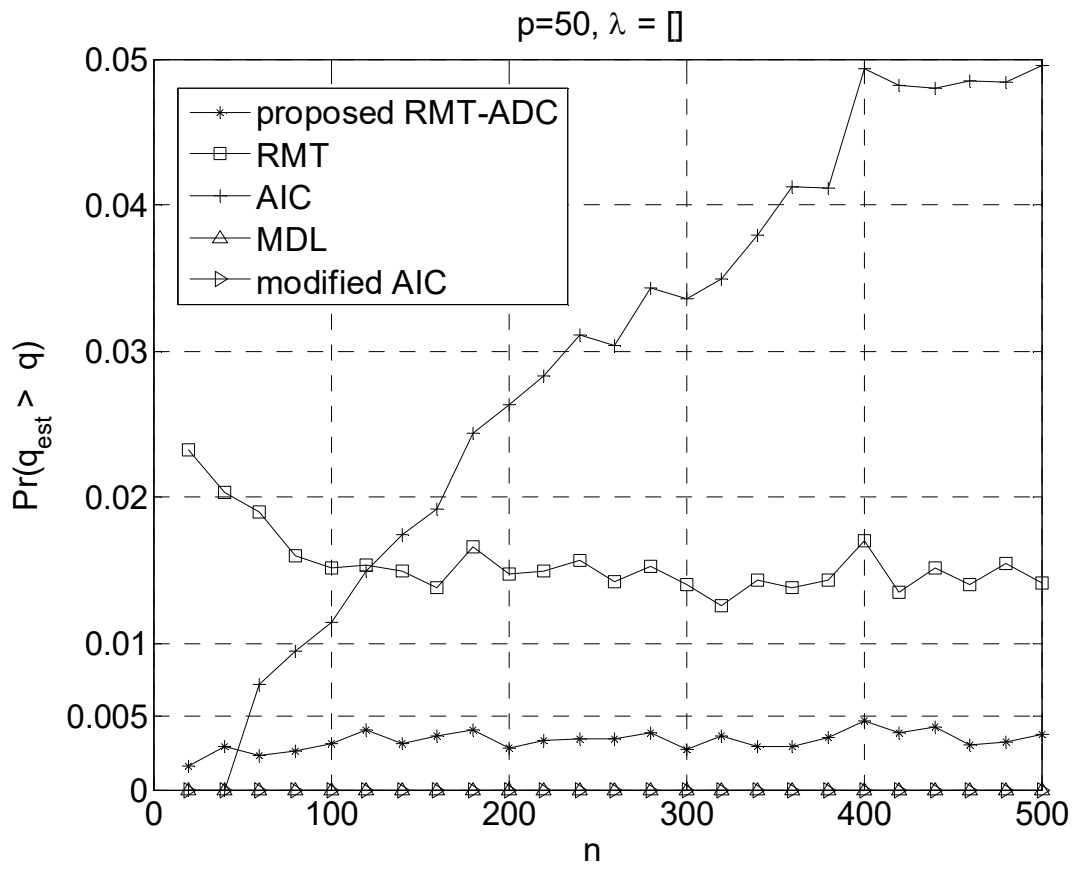

(a)

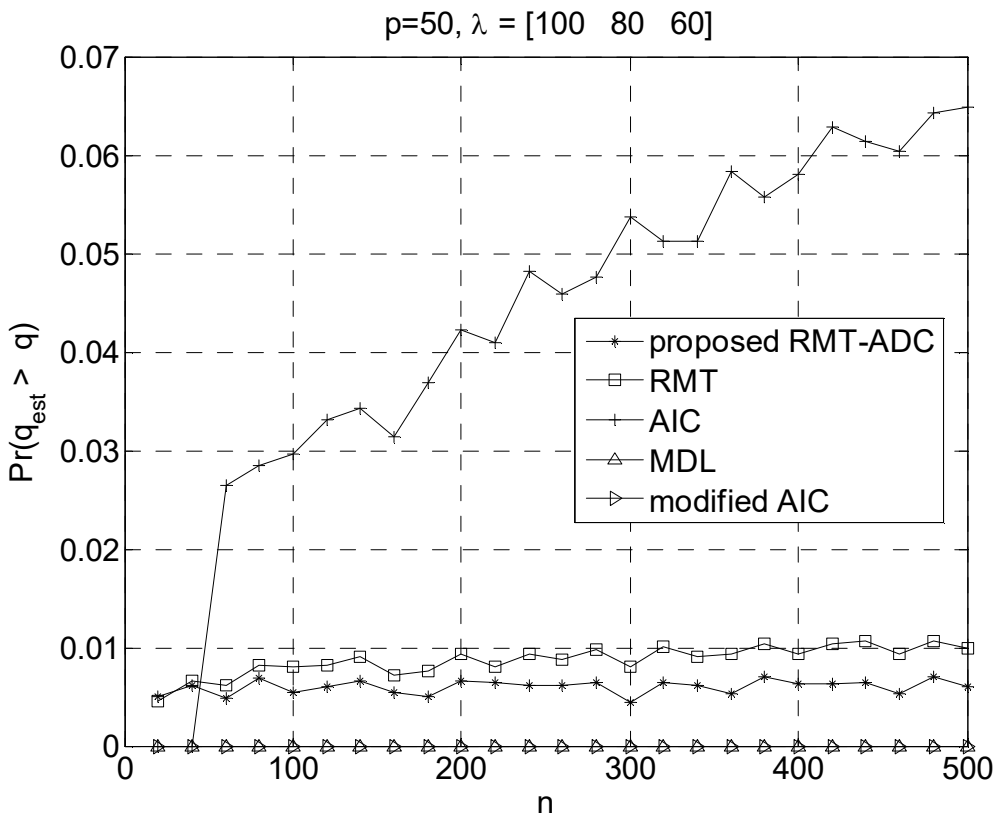

(b)

Fig. 9. Over-estimation probability as a function of sample size $n$ when $p = 50$ : (a) no signal with $\lambda = []$; (b) three strong signals with $\lambda = [100, 80, 60]$ .

Secondly, we examine the effect of various sample size $n$ on the detection performance of various signal number estimators for the case when there are multiple signals. Fig. 10 (a) and (b), respectively, show the misdetection probability and the over-estimation probability of various algorithms as a function of sample size $n$ for the case when there are eleven signals with $\lambda = [12, 10, 9, 8, 7, 7, 6, 6, 5, 4, 2.5]$ when the system size $p = 50$ .

As can be seen from Fig. 10, the proposed RMT-ADC estimator has much better detection performance (with an improvement up to 20%) than the RMT estimator for small to moderate sample size $n$. This is because the RMT-ADC estimator can successfully detect the signal eigenvalue immersed in the bias term among eigenvalues. Moreover, as predicted by the theoretical analysis, the over-detection probability of the RMT-ADC estimator is around the pre-fixed value $\alpha = 0.005$ for all sample size. In addition, though the MDL estimator and the modified AIC estimator have nearly zero over-detection probability, they have much larger under-detection probability than the proposed RMT-ADC estimator. Furthermore, though the AIC estimator has better detection performance than the RMT-ADC estimator for relatively small sample size $n$, it has non-negligible over-estimation probability which is about 10%.

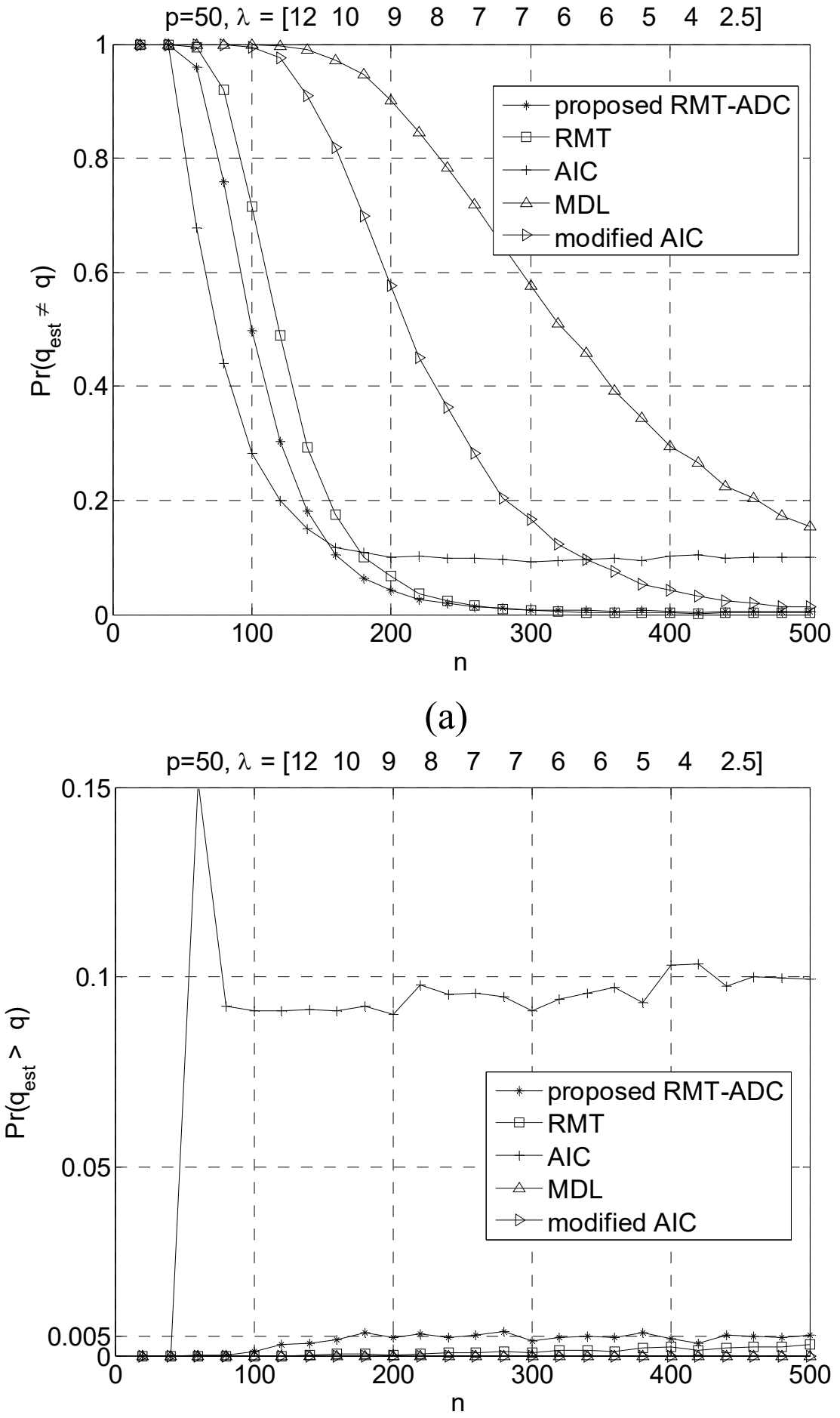

(a)

(b)

Fig. 10. Comparison of (a) misdetection probability and (b) over-estimation probability of various algorithms as a function of sample size $n$ for the case when there are eleven signals with $\lambda = [12, 10, 9, 8, 7, 7, 6, 6, 5, 4, 2.5]$.

## V. Conclusion

As a well-known estimator based on the random matrix theory, the RMT estimator estimates the number of signals via sequentially testing the likelihood of a sample eigenvalue as arising from a signal or from noise. However, the RMT estimator tends to under-estimate the number of signals as signals will be immersed in the non-negligible bias term among eigenvalues for finite system size and sample size. In order to overcome the higher down-estimation probability of the RMT estimator, we have proposed a RMT-ADC estimator with adaptive decision criterion by incorporating the bias term among eigenvalues into the decision criteria of the RMT estimator.

Firstly, we have analyzed the effect of the bias term in (13) among eigenvalues on the detection performance of the RMT estimator. Specifically, we have derived the analytical formulas for both the increased under-estimation probability and the decreased over-estimation probability of the RMT estimator incurred by the bias term among eigenvalues. Secondly, in order to overcome the higher under-estimation probability of the RMT estimator, we have derived both the increased under-estimation probability and the decreased over-estimation probability of the RMT estimator incurred by the bias term among eigenvalues. Based on these results, the RMT-ADC estimator can adaptively determine whether the noise variance should be estimated under the assumption that the eigenvalue being tested is arising from a signal or from noise. Based on these results, the RMT-ADC estimator can adaptively select the decision criterion from (18) or (54), and can also determine whether the bias term should be incorporated into the selected decision criterion or not. Finally, we have shown by extensive simulation results that the proposed RMT-ADC estimator has much better detection performance than the existing estimators including the RMT estimator, the classic AIC and MDL estimators, and the modified AIC estimator.